 \documentclass[acmlarge,screen]{acmart}

\AtBeginDocument{%
  \providecommand\BibTeX{{%
    \normalfont B\kern-0.5em{\scshape i\kern-0.25em b}\kern-0.8em\TeX}}}

\setcopyright{rightsretained}
\copyrightyear{2021}
\acmYear{2021}
\acmConference[MobileHCI '21]{23rd International Conference on Mobile Human-Computer Interaction}{September 27-October 1, 2021}{Toulouse & Virtual, France}
\acmBooktitle{23rd International Conference on Mobile Human-Computer Interaction (MobileHCI '21), September 27-October 1, 2021, Toulouse & Virtual, France}
\acmDOI{10.1145/3447526.3472056}
\acmISBN{978-1-4503-8328-8/21/09}

\acmConference[Mobile HCI '21]{Mobile HCI '21: ACM International Conference on Mobile Human-Computer Interaction}{Sept 27 - Oct 01, 2021}{Toulouse, France}
\acmBooktitle{Mobile HCI '21: ACM International Conference on Mobile Human-Computer Interaction,
  Sept 27 - Oct 01, 2021, Toulouse, France}
\acmPrice{15.00}
\acmISBN{978-1-4503-XXXX-X/18/06}

\begin{document}

\title{Using Computer Simulations to Investigate the Potential \newline Performance of ‘A to B’ Routing Systems for People with Mobility Impairments}


\author{Reuben Kirkham}
\affiliation{%
  \institution{Monash University}
  \city{Melbourne}
  \country{Australia}}
\email{reuben.kirkham@monash.edu}

\author{Benjamin Tannert}
\affiliation{%
  \institution{Hochschule Bremen, City University of Applied Sciences}
  \city{Bremen}
  \country{Germany}
  \email{benjamin.tannert@hs-bremen.de}
}

\renewcommand{\shortauthors}{Kirkham and Tannert}

\begin{abstract}
Navigating from ‘A to B’ remains a serious problem for many people with mobility impairments, due to the need to avoid accessibility barriers. Yet there is currently no effective routing tool that is regularly used by people with disabilities in order to effectively avoid accessibility barriers in the built environment. To explore what is required to produce an effective routing tool, we have conducted Monte-Carlo simulations, simulating over 460 million journeys. This work illustrates the need to focus on barrier minimization, instead of barrier avoidance, due to the limitations of what can be achieved by any accessibility documentation tool. We also make a substantial contribution to the concern of meaningful performance metrics for activity recognition, illustrating how simulations can operate as useful real-world performance metrics for information sources utilized by navigation systems.
\end{abstract}



\maketitle


\section{Introduction}
Accessibility in the built environment remains a perennial problem for the hundreds of millions of people worldwide who have mobility impairments \cite{Organization2011World}. A particularly pertinent challenge is how to navigate from ‘A’ to ‘B’ in a reasonable time, whilst avoiding accessibility barriers (e.g. \cite{Ding2007Design,Wu2020mobile-based,Kamaldin2019Smartbfa:,Volkel2008Mobility,Harriehausen-Muhlbauer2016WheelScout-Barrier-Free}), i.e. ‘barrier avoidance’. One difficulty is the present lack of effective documentation of the built environment, meaning that the existence of accessibility barriers is not known in advance, even for permanent barriers that are often a longstanding feature of the local landscape. Automated tools have limited accuracy even when focused on identifying well-defined accessibility barriers, whilst the impact of a given accessibility barrier can vary greatly depending on an individual’s disability and individual circumstances \cite{Gupta2020Towards,Kirkham2017WheelieMap:}. This means that outside of areas that have already been subject to expert documentation (which is prohibitively expensive in most circumstances), there is insufficient information available to effectively navigate the built environment in a reliable fashion. The wider implication of this is to have a considerable negative impact on the lives of many people with mobility impairments \cite{Bromley2007City,Hara2016design}, and thus limiting their inclusion in wider society.

There has been an increasing amount of research aimed at automatically (or sometimes semi-automatically) documenting the built environment (e.g. \cite{Gani2019Smart,Hara2014Tohme:,Iwasawa2015Road,Kirkham2017WheelieMap:,Mascetti2020SmartWheels:,Mourcou2013Wegoto:,Saha2019Project}). However, the recognition performance of these accessibility documentation systems has not been directly connected with the navigation task that they are supposed to ultimately assist with. At the same time, drafting appropriate descriptive performance metrics is challenging, as the required documentation performance depends on \textit{(i)} the navigation needs of the end user (which can widely vary depending on their disability and personal circumstances), \textit{(ii)} the degree of inaccessibility in the environment (which again, is person specific) and\textit{ (iii)} the topology and layout of the built environment itself. 

We present and apply a (suitably general) simulation framework that enables us to connect the recognition performance of an \textbf{accessibility documentation } system to the performance of an ‘\textbf{A’ to ‘B’ navigation} system that relies on the data generated by it. This means our work is the first that uses simulations as performance metrics for a human activity recognition system. Our bespoke simulation framework was developed in order to enable the simulation of hundreds of millions of journeys in a reasonable time and itself represents an important contribution to accessibility documentation. By running a wide range of Monte-Carlo simulations using our framework (over 460 million journeys in total), we provide answers to important strategic questions which in turn can help shape the important research agenda of accessibility documentation. We find that even a perfect accessibility documentation tool can only offer a reasonable experience in a city that is already largely accessible. Our wider results suggest a different ‘end game’ for accessibility documentation tools: rather than focusing on barrier avoidance, the emphasis should instead be on minimization and accepting the inevitable imperfection of these navigation systems. We also explain how our simulation approach could support this process, and thus provide candid information to end users about the likelihood a route will be inaccessible, thereby providing meaningful feedback to end users.  Finally, we explain how our simulation approach could be adapted for a wide range of navigation scenarios, including navigation scenarios that do not involve people with mobility impairments.  

\section{Background}
\subsection{The ‘A to B’ problem and its implications for people with mobility impairments}
The ‘A \textbf{to} B’ routing problem (sometimes called the sidewalk accessibility problem) involves finding the best route in the built environment to undertake a journey from one location (‘A’) to another (‘B’). It is distinct from the ‘\textbf{within A/B}’ problem which addresses the scenario as to whether a given location itself is accessible (as opposed to an accessible route to getting there in the first place). For many people with disabilities, pre-existing navigation tools are often inadequate, because they do not reliably accurately provide routes that are accessible to the end user \cite{Gupta2020Towards,Tannert2019Analyzing}. In the real world, accessibility barriers often permeate the built environment, with a recent study reporting averages of over 1 accessibility problem for every 100m travelled in some cities \cite{Froehlich2020Towards}, whilst another investigation showed one US city had 80\% of curbs that were not ADA compliant \cite{Bagenstos2020Towards}. Providing effective and accurate navigation tools for people with mobility impairments is thus an important yet difficult challenge \cite{Froehlich2019Grand}, given the large number of pre-existing barriers.

For people with mobility impairments, taking any journey can be burdensome, especially if attempting to do so in a reasonable time, with strategies aimed at somewhat addressing this including ‘reccies’ to avoid barriers \cite{Hara2016design}, or avoiding travelling as a pedestrian where possible \cite{Bromley2007City}. The overall effect is to \textit{“depriv[e] disabled individuals of social opportunities (because they cannot visit friends, family, and places of entertainment or recreation), and [often cause] serious physical injuries when wheelchair users tip over obstacles or are hit by cars when forced to travel in the streets”} \cite{Bagenstos2020Towards}.  Other examples include longer term health damage due to taking suboptimal routes, e.g. to manual wheelchair users by way of cause damaging whole-body vibration \cite{Bowtell2015Assessing,Rice2018Quality}, or in using inappropriately designed ramps that require undue amounts of force to access \cite{Velho2016effect}. At the same time the range of barriers (and thus implications) is diverse, with many different types of concerns being overlooked or underemphasized in existing studies \cite{Gupta2020Towards}. This leads to two types of concerns in respect of accessible navigation: (i) barrier avoidance (aimed at ‘barrier-free’ routes, e.g. as in \cite{Harriehausen-Muhlbauer2016WheelScout-Barrier-Free,Kamaldin2019Smartbfa:,Wu2020mobile-based})  and (ii) a ‘quality aware’ approach \cite{Siriaraya2020Beyond}, which simply seeks to minimize the number of barriers encountered.

\subsection{Accessibility Documentation Systems}
An effective navigation tool needs to be informed by accurate information about the accessibility of the built environment. This is easier said than done: indeed, even defining accessibility can be challenging \cite{Church2003Measuring,Saha2021Urban} and there is a wide range of different types of physical barriers (whose effects in turn depend on an individual's specific disability). Nevertheless, there are various types of accessibility documentation systems that have been designed in an effort to provide this type of information, typically by identifying specific types of physical features. Some of these approaches are human-driven and rely on individuals to collect the relevant information. One of these approaches is to use experts, however, whilst sufficiently accurate, this approach is expensive and results in very limited coverage due to these exercises being \textit{“laborious and time consuming”} \cite{Froehlich2019Grand}, meaning that most cities do not actually have anything approaching an accurate map of accessibility barriers \cite{Bagenstos2020Towards,Bromley2007City}. Another strategy that has been tried is ‘geocrowdsourcing’ \cite{Rice2018Quality} yet this has encountered significant problems in respect of their ability to accurately document barriers \cite{Rice2018Quality} and has also lacked effective engagement and thus coverage \cite{Froehlich2019Grand,Mascetti2020SmartWheels:}.  

Given the limitations of human-driven approaches, a growing body of accessibility documentation research has focused on using automated or semi-automated approaches (e.g. \cite{Hara2014Tohme:,Kirkham2017WheelieMap:,Saha2019Project,Weld2019Deep}), with \cite{Lange2021} providing a body of principles for designing appropriate automated accessibility documentation systems. To date, the most successful automated example is based on Project Sidewalk data \cite{Saha2019Project}, which used computer vision (and RESNET) to achieve a f-score of 0.85 in a best case scenario \cite{Weld2019Deep} (albeit with some variation across classes, with weaker performance on ‘surface problems’). Taking into account the difficulty of documentation this is a strong recognition performance, especially given the variety of forms that physical barriers can take, and the disagreement amongst people with mobility impairments as to the real-world impact of given barriers on accessibility \cite{Kirkham2017WheelieMap:}. 

We also note that there are numerous papers that claim to offer a viable system for using inertial sensors mounted on mobility aids (e.g. wheelchairs) to automatically measure the accessibility of the built environment \cite{Iwasawa2015Toward,Iwasawa2012Life-logging,Iwasawa2016Combining,Kurauchi2019Barrier,Yairi2019Estimating}. Unfortunately, the evaluations of these initial sensor based systems are typically problematic in two different ways: they are not naturalistic (or ecologically valid) which means that the study’s findings do not translate to realistic activity recognition problems \cite{Poppe2007Evaluating} and they fail to use the appropriate leave-one-out metric of evaluation \cite{Hammerla2015Let's} (in some cases, participants were left out, but they all followed the same route, which is a different variant of the same mistake). The recent work of \cite{Mascetti2020SmartWheels:} showed that there is a stark difference in reported performance between a leave one out approach and a k-fold evaluation in the context of obstacle detection (with a f-score 0.616 for leave one out and 0.84 for the inappropriate k-fold evaluation). Accordingly, whilst there are systems that purport to have an f-score above 0.9, these systems have not been shown to have a realistic chance of working effectively in the real world. 

\subsection{The Fair and Meaningful Metrics for Activity Recognition Problem}
This paper is also a different take on the meaningful performance metrics for activity recognition problem, as existing documentation tools are often activity recognition systems (where the ‘activity’ is some sort of interaction with a potential barrier). At present, we are unaware of any cases where the performance of these automatic documentation systems has been directly connected to real navigation tasks. There is an important gap, in that there are no current performance metrics that convert the documentation performance into navigation performance. More generally, the determination of appropriate performance metrics is a matter of active investigation in respect of activity recognition \cite{Hammerla2015Let's,Ward2011Performance}. Our work accordingly examines how we can translate documentation performance into navigation performance, using computer simulations to do so. 

At the same time, the meaningful metrics problem is also of concern in respect of ‘fair AI’ more generally. Whilst Fair AI has primarily focused on the protected attributes of gender and ethnicity, the need to ensure that people with disabilities has been mostly overlooked, being \textit{“largely omitted from the AI-bias conversation”} \cite{Whittaker2019Disability}. In particular, there is a real need to ensure that the diverse range of considerations in respect of disability are considered: the variety of disabilities makes training a AI system based on machine learning more challenging \cite{Morris2020AI}, whilst meaning that inherent biases are less likely to be detected. In the context of accessibility documentation tools, this might arise by inadvertently taking a narrow view of disability, and thus excluding subtle instances of accessibility barriers. If an accessibility documentation system is reliant on using the motion of a disabled person, then a particular impairment (e.g. an unusual gait) may also reduce recognition performance \cite{Morris2020AI}, and thus the reliability of documentation. Understanding the real world implications of a given systems recognition performance is thus an important consideration for ensuring people with disabilities are not disadvantaged by the use of AI systems, including to document inaccessibility in the built environment: indeed it is very similar to design engineering, which has also been recently advanced for more effective technology design for people with disabilities by conducting simulations \cite{Kristensson2020Design}. 

\section{Connecting Documentation and Navigation}
\subsection{An ‘End-to-End’ system for a diverse range of concerns}
An end-to-end documentation-routing system involves three different interrelated stages:
\begin{enumerate}
    \item The documentation of the built environment and whether a given path segment is accessible.
    \item The design of a routing algorithm that proposed the most appropriate routes (and indicates its confidence in its own predictions)
    \item An interface that sits on top of the routing algorithm so the user can apply the results to navigate from ‘A to B’
\end{enumerate}

The difficulty is that these steps have been siloed from one another, when in reality they are closely interconnected. A routing algorithm can only operate on the accessibility information it is given. And the perceived performance of a navigation ‘interface’ in turn depends on how good the results provided by its underlying algorithm (as well as how that information is presented). At the same time, better documentation can also come from the reporting of system errors and mistakes by end users of navigation systems: but only if they are satisfied enough with a navigation system to engage with it. The result (\textbf{Figure 1}) is that these navigation and documentation challenges are heavily interconnected: nevertheless, in practice, they are studied in silos that need to be brought together. Notably, user studies deal with only the interface itself, which is several layers removed from the underlying accessibility documentation system. 

\begin{figure}[h]
  \centering
  \includegraphics[width=\linewidth]{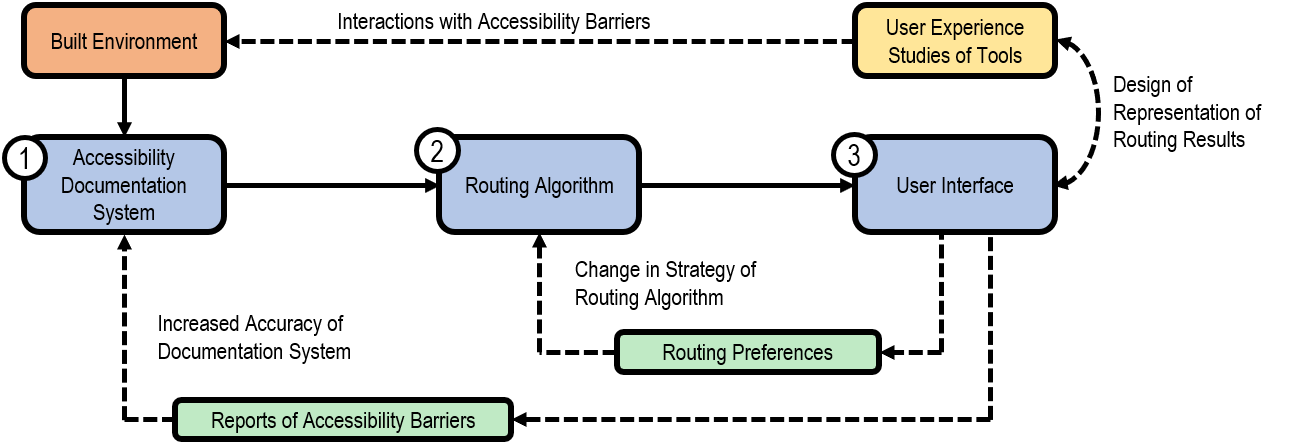}
  \caption{Illustration of feedback loops that influence the design and performance of an accessibility documentation tool.}
  \Description{This illustrates the relationship between each element of the feedback loop in the design of a navigation tool for people with mobility impairments. The process starts with the built environment, which is observed by the 'accessibility documentation system'. The data from the 'accessibility documentation system' is then used by a 'routing algorithm' to propose routes. The results of the 'routing algorithm' are then presented to the user in the 'user interface'. The 'user interface' in turn can provide reports of inaccessibility (which feed into the documentation system, improving its performance), as well as allowing the selection of 'routing preferences', which influence the decision making of the 'routing algorithm'. At the same time, this is difficult to directly study - a user experience study only sees the interface in a given built environment, rather than the wider eco-system.}
\end{figure}

Adding to that complexity is the inherently person-specific nature of the accessibility documentation problem: different people have different disabilities and personal circumstances, and thus different needs and expectations. Yet as explained in \cite{Gupta2020Towards}, there is a tendency for navigation support approaches to be focused on one disability group, even though there are a wide range of impairments that a fully inclusive end-to-end navigation system needs to support. Although existing navigation systems often assume otherwise, different disabled people can have different objectives: one user might be interested in purely avoiding barriers (and thus is concerned with ‘barrier avoidance’ approach), whereas another might be satisfied with simply encountering them appreciably less often (i.e. the ‘quality aware’ routing approach), as these barriers are discomforts, rather than hard-edged dangers or obstacles. As such, any framework needs to be suitably abstract and flexible to deal with a broad range of navigation concerns and objectives: there is no fixed optimal solution given the diversity amongst disabled people. At the same time, the output representation also has to be optimized to suit the preferences and needs of end users, but determining how to best do this is tricky if studies are confounded by errors in the underlying accessibility documentation tool.

A real world understanding of what an accessibility documentation tool can accomplish is needed to inform the most optimum approach for designing navigation tools. In effect, this is a ‘design engineering’ [24] problem, wherein simulations can help determine the most appropriate form of system. The distinction is that we are using these simulations to create real-world performance metrics, that provide a clear illustration of the navigation performance of a system based on a documentation tool with a given recognition performance. These metrics in turn describe the bounds on what can be accomplished with a navigation system in the real world, thus helping to determine the best strategy for using the information available to it based on an imperfect documentation of the built environment. In turn, these answers inform expectations for any feedback loop that might improve performance. 
\subsection{Using Monte-Carlo Simulation to Connect Routing and Documentation}
To connect the performance of an accessibility documentation system to a navigation system that relies upon the information provided by it, we conducted extensive Monte-Carlo simulations, with over 460 million runs (or simulated journeys) in total. Our simulations were constructed to focus on the cases where a person is seeking to engage in \textbf{barrier avoidance}, as this is the stated goal of a substantial proportion of work that focusses on accessible navigation, with a view towards investigating the viability of this approach. A \textbf{map }contains a \textbf{set of edges (or path segments)} connected at \textbf{nodes (or junctions)}: a proportion of these edges/path segments will have accessibility barriers (\textbf{Figure 2}). The\textbf{ Inaccessibility Rate} is the proportion of edges that are inaccessible in our simulation for our hypothetical person with a mobility impairment. Our starting point is that an accessibility documentation system has both a \textbf{True Positive Rate (TPR)} which reflects the proportion of edges containing accessibility barriers (for that person) that it accurately detects, and a \textbf{True Negative Rate (TNR)}, which indicates the proportion of accessible edges it correctly labels as being such. The Inaccessible Rate, the TNR and the TPR are all weighted (linearly) by distance, so that longer segments are more likely to have an accessibility barrier. On a \textbf{map}, this translates into marking a specific list of edges as being inaccessible from the \textbf{perspective} of the documentation tool – unless the tool has a perfect TPR and TNR, then this perspective will sometimes be inaccurate, potentially leading to the person in question being routed inappropriately (see \textbf{Figure 3} for examples of this and the different types of difficulties it can cause). A navigation tool will then automatically choose the shortest route based on the barriers identified by the documentation tool. 

\begin{figure}[h]
  \centering
  \includegraphics[width=0.4\linewidth]{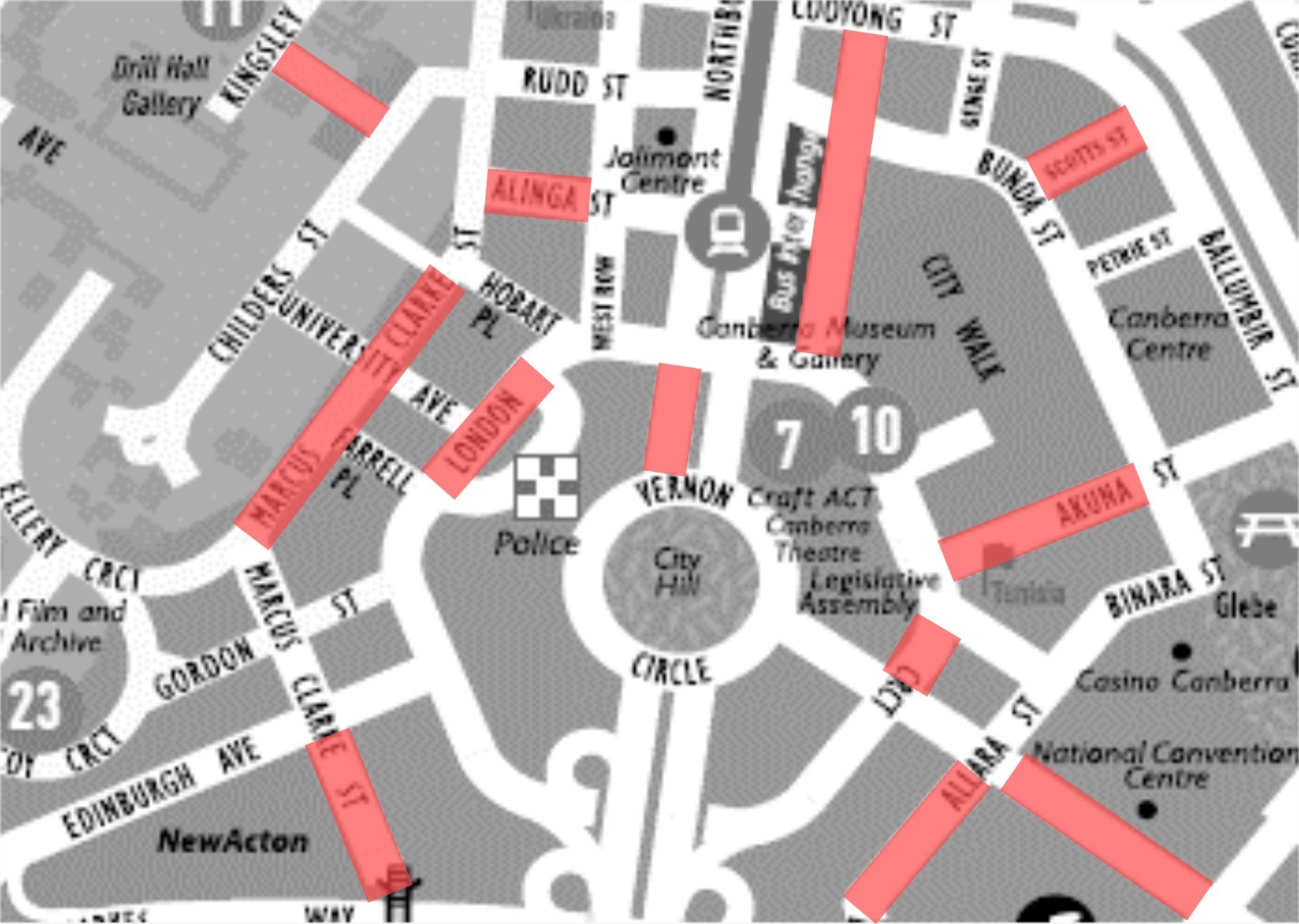}
  \caption{Illustration of a map, where inaccessible edges or path segments are marked in Red.}
  \Description{}
\end{figure}
\begin{figure}[h]
  \centering
  \includegraphics[width=\linewidth]{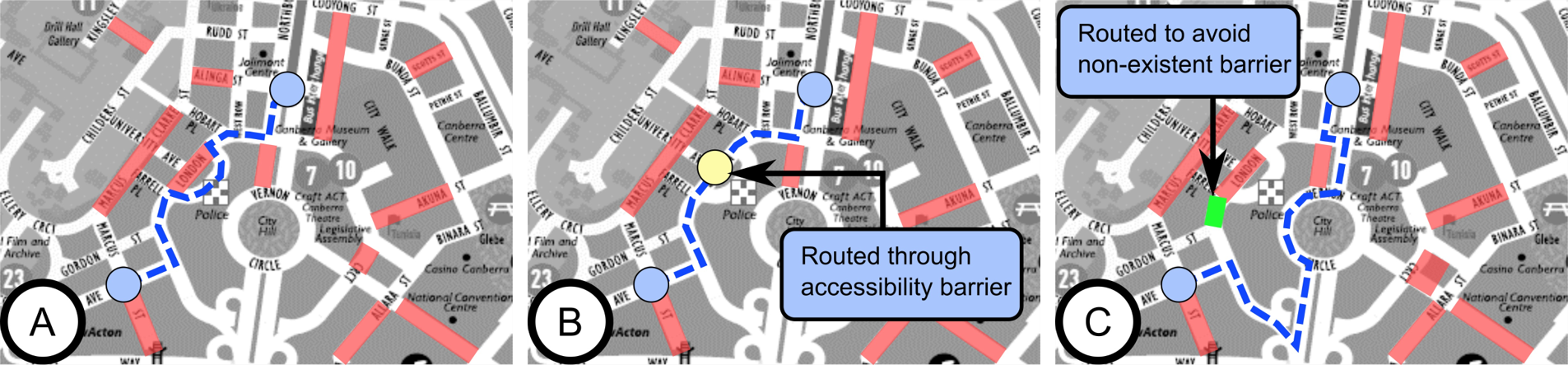}
  \caption{Illustration of different routing scenarios and the consequences of different types of documentation error. A shows the most optimal route in the real world. B illustrates a shorter route being taken that encounters an undocumented barrier (a false negative), whilst C illustrates a longer route being taken due to an attempt to avoid a non-existent accessibility barrier. }
  \Description{}
\end{figure}

To enable us to simulate such a large number of journeys, we structured our simulations with a bespoke design that enabled computational efficiency (\textbf{Figure 4}), developing our own novel approach for doing this based on ablation and pre-computation. After extracting regions from OpenStreetMap, we then generated a stratified sample (by distance) of different location pairs (‘A’ and ‘B’) up to a ‘crow flies’ distance of 1.2km (very short routes less than 300m were excluded): we selected 60 journeys in total. For each journey, we then \textbf{pre-computed }a list of all possible routes (and the distance) between ‘A’ to ‘B’ up to a limit of 1.5km total edge length using the CPU’s on an HPC cluster. Each simulated journey involved generating a list of inaccessible segments on the map. We then generated a list of segments actually determined to be inaccessible by the documentation tool (based on the information available to it). The route to be followed was then computed using an ablation approach on a GPU (with a speed increase of around 1000 compared to a single CPU core). This ablation approach involves identifying all the edges perceived as being inaccessible and deleting them from the network, which computationally means deleting the precomputed routes that contain one or more such edges and selecting the shortest remaining route, if it exists\footnote{If no route exists, then the system would report it is not possible to complete the journey.}  (the deletion is a matrix operation that can be done rapidly on the GPU). Then the relevant performance metrics, including whether an accessibility barrier was counted and if the system sent a person on an unduly long route were computed based on the selected route, thus enabling a report on the different types of errors and in turn, a cost benefit analysis to be conducted.   
\begin{figure}[h]
  \centering
  \includegraphics[width=\linewidth]{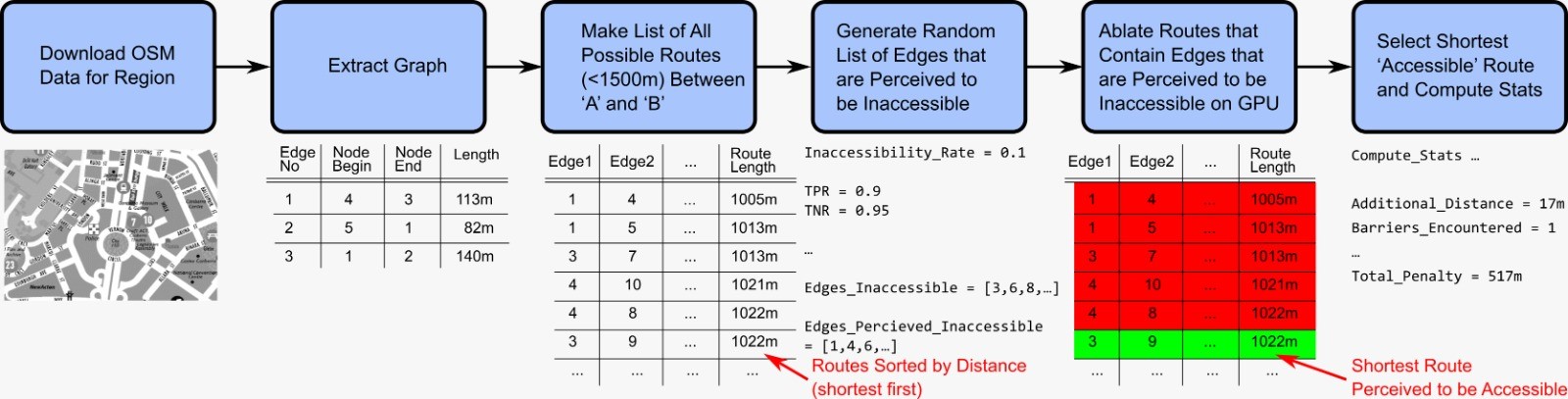}
  \caption{ Overview of Process for Simulating Journeys. The list of all possible routes is generated by CPU’s on a high-memory HPC node (>100GB), whilst the ablation step is computed on a NVidia P100 GPU. Each Route List is used 7.68 Million times in our simulations, thus making the simulations computable in a reasonable amount of time.}
  \Description{}
\end{figure}

\subsection{Our Simulations in More Detail}
\subsubsection{Datasets}
The performance of a routing algorithm depends partially on the geometry or structure of the city. This is because the consequences of an error will differ depending on factors such as the time taken to ‘go around’, as well as the availability of different options. We therefore extracted the geometries of three different cities (Canberra, Prato and Seattle) from Open Street Map. Each city was chosen due to having substantially different geometries, both in terms of the average number of edges connected to each node, and the differing volume of street coverage.\footnote{To give some examples, according to OSMnx \cite{Boeing2017OSMnx:}, the regions we selected of Canberra, Prato and Seattle have average number of nodes per intersection of 3.30, 2.84 and 3.71 respectively, whilst the edge density (per km2) was respectively 20512, 18454 and 28554.}  These geometries included all pedestrian pathways available in the region. For our analysis, this geometry was converted into a graph comprising a list of edges, their lengths, and which nodes they are connected to at each junction. 

For each city, we selected 20 routes from A to B (up to 1.2km apart in ‘crow-flies’ distance) between different landmarks in the city, with a view towards creating a representative ‘bag of journeys’ for our simulations. Very short routes (less than 300m) were excluded, given that they are unlikely to be the use case for a navigation tool. This led to a total of 60 journeys in our ‘bag of routes’.  The distance range we selected has the advantage of making our simulations tractable, as well as being in accordance with the types of journeys typically taken by pedestrians, with the great majority of journeys being included in this range (see e.g. \cite{Yang2012Walking} for summary of trip statistics amongst the general population). We chose accessible bathrooms as landmarks for the reasons given in \cite{Tannert2019Analyzing}, namely this is a commonly required trip by people with disabilities. 
\subsubsection{Modelling Distribution of Accessibility Barriers}
The impact of errors of a documentation system have more or less impact depending on how many accessibility barriers there are in the built environment (for a given individual). Our modelling approach is deliberately abstract, in that we are not concerned about the specific form of barrier (given the variety of different barriers and the different views that each disabled person can have in respect of their importance). Instead, we focus on there being an \textbf{inaccessibility rate} which is unique to a given individual in a given city, and represents the likelihood of that person encountering a barrier. We further assume that accessibility barriers are randomly distributed within a given city and model different probabilities of an \textbf{average length path segment} (i.e. a graph edge) being inaccessible. We assume no priors in distributing accessibility barriers: given the subjective nature of accessibility barriers to an individual and the variety of factors that influence barriers (e.g. level of maintenance, the age of infrastructure and so forth), there were no specific priors we could use, and attempting to do so (e.g. by making major roads less likely to have accessibility barriers) would have introduced a further parameter into our simulation, thus reducing its tractability. Moreover, a city that has launched an extensive enough improvement campaign for barriers to be less randomly distributed will doubtless have used the ‘expert’ documentation approach to document barriers to begin with, yet it is highly unusual for this to be done (with the automated and crowdsourced documentation systems which we model in this work being aimed at overcoming the cost of expert documentation).

In our approach, the probability of a segment being inaccessible was linearly weighted by segment length – for instance, if a segment was twice the length of the average one, then it would be twice as likely to be inaccessible. This makes sense for most types of accessibility barriers, as they are uniformly distributed (e.g. a longer street is more likely to have trip hazards due to surface wear and tear). For our simulations, our\textbf{ inaccessibility rate} ranged from 0 to 0.3 (inclusive), with increments of 0.02 (yielding 16 different values). We excluded values above 0.3 as our pilot simulations indicating that these scores would produce very few accessible routes, thus making a navigation tool infeasible.  
\subsubsection{Modelling Performance of Accessibility Documentation System}
As explained in \cite{Tannert2019Analyzing} (see also \textbf{Figure 3}), there are two main types of documentation error that can be made by an accessibility documentation system. The first is to state that a street segment (i.e. an edge) is accessible when it is not (i.e. a \textbf{false positive}). The second is to mark an accessible edge as incorrectly being inaccessible (i.e. a \textbf{false negative}). For our purposes, this leads to two parameters which describe the performance of our (hypothetical) accessibility documentation systems:

\textbf{True Positive Rate (TPR):} The proportion of inaccessible segments correctly labeled as being such, weighted by distance. In other words, this is the probability of an average length inaccessible segment being correctly labelled. 

\textbf{True Negative Rate (TNR):} The proportion of accessible segments correctly labelled as being such, weighted by distance. In other words, this is the probability of an average length accessible segment being correctly labelled.

As with the inaccessibility rate, both the TPR and TNR are linearly weighted by distance. For both TNR and TPR, the performance of the documentation tool was on the range of 0.7 to 1 (inclusive), with increments of 0.01 (leading to 31 different values for each of them, or 961 combinations of TPR and TNR). This performance range includes the best performing existing systems and takes into account that a system that performed worse that 0.7 under either metric is unlikely to be useful (which we observed when conducing pilot simulations), as well as the fact that there are systems that have results substantially above 0.7 (e.g. \cite{Weld2019Deep})  
\subsubsection{Routing Approach}
We compute the shortest accessible route (i.e. with no reported accessibility barriers) as perceived by the documentation system. If there no accessible route available, the system will report the route as being ‘impassible’. To put this into practice, we compiled a list of all possible routes (with a route being a list of edges) from ‘A’ to “B’ up to a maximum of 1500m route length, and then sorting each route ascending by distance. This list of routes was computed using an HPC cluster, using 96GB RAM and 11 CPU cores for each journey. \footnote{As the cluster made available 65 CPU cores for us, we could run five such jobs at one time. The other CPU’s were used to support the GPU’s. }  

Computing the perception of the system is done in two stages. The first is generating the ‘ground truth’ of where the accessibility barriers really are: a list of edges is randomly generated in line with the \textbf{inaccessibility rate}. The second is randomization of errors made by the documentation system based on its performance, where the relevant proportions of errors are applied to the ground truth in line with the \textbf{TPR} and \textbf{TNR}. After both steps are complete, the navigation system has a list of edges which it perceives contain accessibility barriers. As these are Monte-Carlo simulations, this step is repeated 500 times for each combination of TPR, TNR and inaccessibility rate, meaning 7.688 million journeys are simulated for each pair of locations ‘A and B’ (we have 60 pairs in total). 

The system then selects the optimum route. We do this by ablation\footnote{This is ablation, because it operates by creating a list of edges that are deemed inaccessible, and in effect, deleting them from the network.}, where we identify the subset of all possible routes that do not contain an inaccessible edge (from the perspective of the navigation system): i.e. the routes that are deemed accessible. The shortest accessible route is selected. This is done by way of a matrix operation on a Nvidia P100 GPU within our HPC cluster, which allowed us to use 8 GPU’s at the same time.  
\subsubsection{Performance Measures}
Measuring performance requires the identification of the mistakes that can be made by a routing tool. There are three main types of navigation error (\textbf{Table 1}) that are of concern: (i) providing unnecessarily longer routes, (ii) heading down a route segment that contains a barrier and (iii) reporting that there is no accessible route when one exists.
\begin{table*}
  \caption{Types of Navigation Error that can be made by a navigation system and the relevant implications of each type of Navigation Error.}
  \label{tab:errortypes}
  \begin{tabular}{|p{2cm}|p{6cm}|p{6cm}|}
    \toprule
    Error Type &Description	&Real world cost\\ 
    \midrule
   A	&Reporting a journey is ‘impassible’ when an accessible route exists.	&The journey is made by alternative transport (e.g. a Taxi) or not made at all by the person with a disability.  \\ \midrule
    B	&Providing an unnecessarily longer route.	&The person has to travel this additional distance. Alternatively, they choose to use alternative transport (e.g. a Taxi) or not to make the journey at all. \\ \midrule
C	&Providing a journey that contains an inaccessible segment. 	&The person encounters an accessibility barrier. At the least, they have to reroute around the barrier. Depending on the nature of the barrier, there could be wider consequences (e.g. if the barrier in question is an unexpected trip hazard). \\
    \bottomrule
  \end{tabular}
\end{table*}
In line with \cite{Tannert2019Analyzing} we calculate performance of a tool on a given route by using a distance measure, which is the distance of the route selected, with a further 500m penalty for each accessibility barrier encountered. This estimate is based on a realistic amount of time to select and travel an alternative route that does not contain a barrier. For example, a route of 1100m length where there are two accessibility barriers is counted as 2100m (actual distance of 1100m plus 2 accessibility barriers of a 500m penalty). Our results are reported in relative terms, with three comparators in mind:
\begin{enumerate}
\item \textbf{Perfect World:} This is the performance relative to a world with no accessibility barriers in it (i.e. an inaccessibility rate equal to 0). It represents the experience of someone without any relevant mobility impairment. 
\item \textbf{Perfect System:} This is a routing algorithm which has a TPR and TNR both equal to 1. It represents the best possible performance of a routing algorithm and always chooses the shortest accessible route (if one exists). 
\item \textbf{Oblivious System:} This system has no knowledge of accessibility barriers and thus assumes every segment is accessible (which is equal to a TNR of 1 and a TPR of 0).  
\end{enumerate}
These comparators are chosen as they relate to practical scenarios. A comparison between a ‘perfect world’ and a ‘perfect system’ places a bound on how useful any accessibility documentation tool can be. At the same time, comparing any system to an ‘oblivious system’ provides a positive indication of the utility of an accessibility documentation system in terms of an improvement relative to the status quo of not having any accessibility information. In what follows, we present an \textbf{exploratory analysis}, which illustrates what an accessibility documentation tool is capable of in respect of differing inaccessibility rates, as well as the TPR and TNR of any given tool. 

\section{Results}
\subsection{What is an Ideal Documentation Tool Capable of?}
We first present our simulation results concerning an ideal documentation tool, which has perfect knowledge of each and every accessibility barrier, as these places a limit of how much benefit can be gained from any documentation tool. \textbf{Figure 5(i)} illustrates the influence of the inaccessibility rate, whilst \textbf{Figure 5(ii)} illustrates the increase in distance required for the remaining routes. As can be seen, a city that has an increase by even a moderate volume of accessibility barriers offers a severely impaired navigation experience. There is also a (relatively small) difference between different cities in our ‘bag of routes’, with our simulation of Prato being the most challenging, and Canberra being somewhat less affected by an increased inaccessibility rate (this difference is limited is despite the fact that these cities have markedly different structures, especially in terms of numbers of intersections per node\footnote{Prato likely performs worse overall due to the relatively reduced number nodes per intersection and its lower street density. Further investigation (beyond the scope of this paper) is needed to indicate the full implications of geometry on the relative (in)accessibility of a city.}). These results clearly justify our focus on routes with a relatively low proportion of accessibility barriers, as with an inaccessibility rate above 0.15, it is unlikely that even a perfect tool will be able to offer a sufficiently useful navigation experience, given the increase in distance and (more markedly) the proportion of routes that are unavailable. 

\begin{figure}[h]
  \centering
  \includegraphics[width=\linewidth]{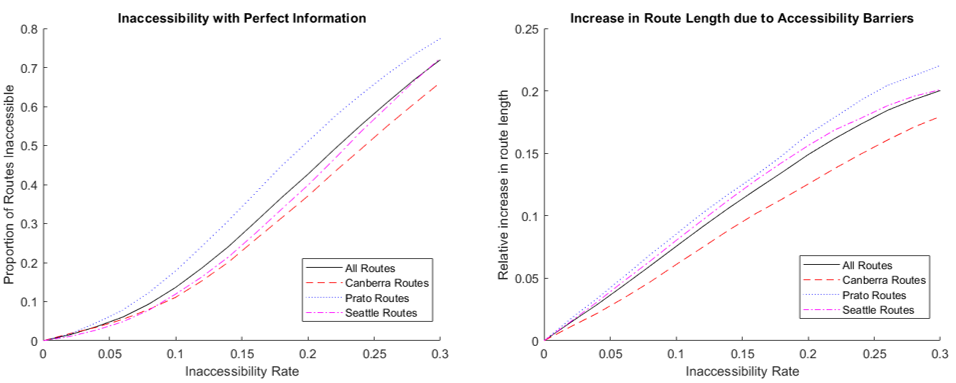}
  \caption{ On the left is Figure 5(i), which illustrates the proportion of routes that are actually impassible, whilst on the right is Figure 5(ii), which illustrates the increase in distance (as a proportion of the route length) for the shortest accessible route. .}
  \Description{}
\end{figure}

\subsection{Accurate and Inaccurate Reporting of Impassible Routes}
We consider that a route is\textbf{ impassible} if there is no possible way to travel from A to B without interacting with at least one accessibility barrier, otherwise we call it\textbf{ navigable}. A perfect tool (where $TPR=TNR=1$) will always accurately report on impassible routes. However, a tool that is inaccurate can wrongly report routes to be navigable, when they are not, or vice versa. \textbf{Figure 6(i) }illustrates the effect of the TNR upon navigable routes that are falsely reported to be impassible in the case where the $TPR=1$. As can be seen, a high TNR is necessary if a large proportion of accessible routes are not to be wrong reported as being impassible. In a realistic scenario of an inaccessibility rate of 0.2, a TNR of 0.95 would still wrongly report nearly 30\% of all accessible routes as being impassible, whilst a TNR of 0.9 would lead to over 50\% of routes being falsely reported as such. Even in environments where there are no accessibility barriers at all (i.e. an inaccessibility rate of zero), a substantial proportion of routes can falsely be reported as being impassible, even with respect to relatively high TNR’s (e.g. at a TNR=0.9, nearly 15\% of routes are falsely reported as being impassible).  

\begin{figure}[h]
  \centering
  \includegraphics[width=\linewidth]{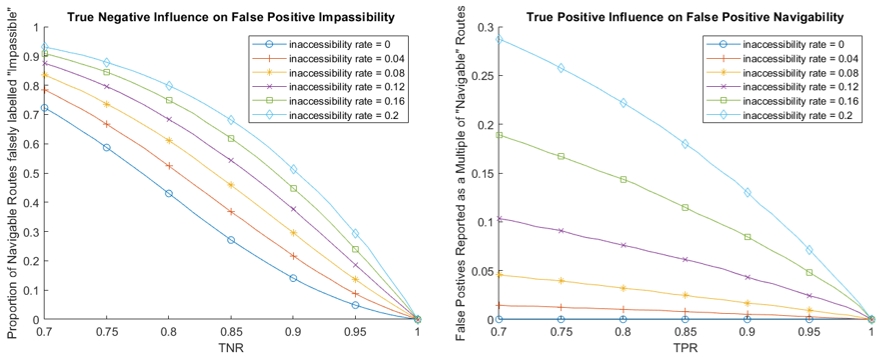}
  \caption{ On the left in Figure 6(i) is the illustration of missing routes performance when the TPR is fixed to 1, with differing TNR values and inaccessibility rates. In Figure 6(ii) is the case where the TNR is fixed to being 1, where we can see the influence of the TPR on routes being falsely reported as being navigable, when they are in fact impassible. }
  \Description{}
\end{figure}

A somewhat lesser problem is a risk of false positives, namely routes are declared to be navigable when in fact they are not. As can be seen in\textbf{ Figure 6(ii)}, a lower TPR can lead to this issue occurring quite often: for instance, at a TPR of 0.85 and an inaccessibility rate of 0.2, this happens nearly 20\% of the time. It is also worth noting that the inaccessibility rate is also a particularly influential factor, with the increase being non-linear, and this occurrence being rare at lower inaccessibility rates, even with a relatively low TPR – for example with an inaccessibility rate of 0.08, and a TPR of 0.7, this still happens less than 10\% of the time.   

\subsection{ Performance of Navigation Tool versus an Ideal Navigation Tool on Navigable Routes}
We now present the performance of a navigation tool for the cases where it reports a navigable route \textbf{and} the reported route is indeed navigable. We do this by applying the distance measure set out in\textbf{ Section 3.3.5}, looking at the relative increase in distance relative to a perfect navigation tool and taking into account the accessibility barriers encountered. Noting that an ideal tool does not encounter accessibility barriers, this is as follows:

\begin{equation*}
score=  \frac{(dist_{tool}+500\times nbarriers_{tool})-dist_{perfect}}{dist_{perfect}}
\end{equation*}
In that formula, $dist_{tool}$ is the distance of the route selected by the routing tool, $dist_{perfect}$ is that of the journey selected by an ideal tool (i.e. where $TNR=TPR=1$) and $nbarriers_{tool}$  is the number of edges with accessibility barriers in the journey proposed by the navigation tool. All distances are in meters. The resulting performance is illustrated by surface plots in \textbf{Figure 7}.\footnote{Note that the plots in Figures 7-9 are fitted surfaces generated using MATLAB \textsuperscript{TM}’s Curve Fitting Toolbox (and fourth degree polynomials).}  It is clear that the higher the inaccessibility rate, the worse the given score as a proportion of distance. Both TNR and TPR have an impact on performance, however lower TNR’s have less of an impact (on routes already reported as being passible), with this being particularly clear when one compares charts with the lower inaccessibility rates. At higher inaccessibility rates, the impact of lower TPR’s (and TNR’s) is particularly pronounced, with many journeys being substantially increased in relative length. It can be seen from\textbf{ Figure 8(i)} that much of this arises from barriers being interacted with, as opposed someone being simply sent on a longer route. 

\begin{figure}[h]
  \centering
  \includegraphics[width=\linewidth]{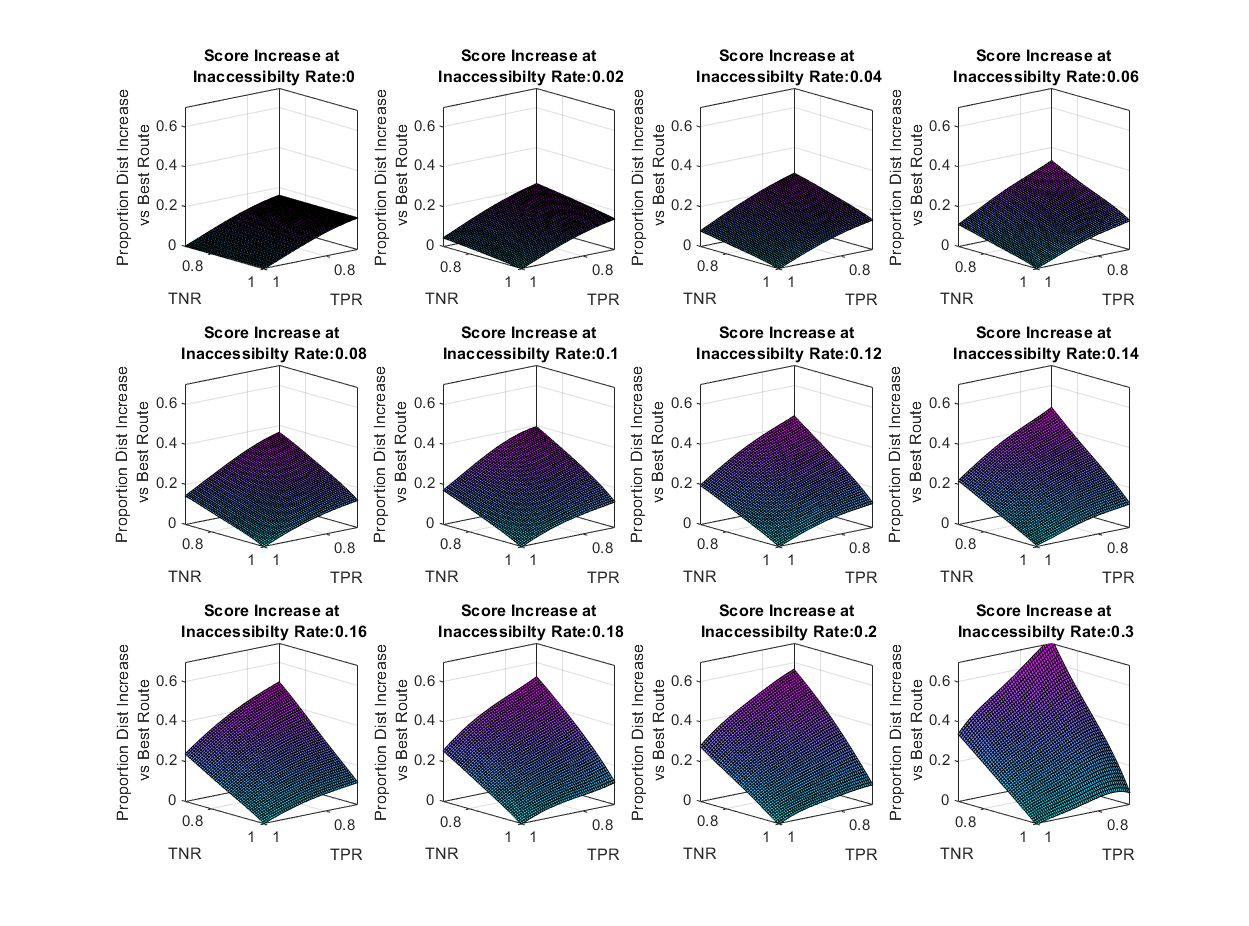}
  \caption{ Relative Score increase relative to a perfect tool in respect of different inaccessibility rates, for tools with different TNR and TPR values. As can be seen, at lower inaccessibility rates, the TNR has a lesser influence than the TPR. Higher inaccessibility rates also have a marked influence on the effect of lower TPR and TNR’s.  }
  \Description{}
\end{figure}
\subsection{Performance of Navigation Tool versus an Oblivious Navigation Tool}
The previous three subsections considered the performance of a navigation tool versus an ideal routing tool. We now consider the performance of a routing tool with different TNR and TPR relative to an oblivious navigation tool (i.e. one with $TNR=1$, but $TPR=0$), for the cases where a navigation tool reported an accessible route (notably an oblivious tool will always report there being an accessible route). As can be seen from \textbf{Figure 8(ii)}, even with relatively moderate inaccessibility rates, e.g. 0.1, the increase in accessibility barriers encountered on an average route is over 1 (i.e. on each journey, one would expect to encounter an accessibility barrier over and above that captured by the navigation tool): by contrast, even with a TPR of 0.7 and an inaccessibility rate of 0.2, one would not necessarily encounter a barrier any more than 50\% of the time (per Figure 8(i)). Thus, even relatively poorly performing tools can have a substantial impact on barrier avoidance and provide a markedly better experience.  A strongly performing tool would avoid barriers most of the time.
It is also important to consider performance by way of distance. We compute a relative score as follows:
\begin{equation*}
score= \frac{(dist_{tool}+500 \times nbarriers_{tool})-(dist_{oblivious}+500 \times nbarriers_{oblivious})}{dist_{best} }
\end{equation*}
All the above have the same meaning as with the previous score, whilst $dist_{oblivious}$ is the distance of the route proposed by the oblivious tool (i.e. $TPR=0$,$TNR=1$), and $nbarriers_{oblivious}$   is the number of barriers encountered. The formula is weighted in proportion to the best possible performance. The result of the exercise is provided in \textbf{Figure 9} (with\textbf{ negative }distance increases being better than an oblivious system). Asides cases where there is a relatively poor TPR and TNR, the oblivious tool is always outperformed, even at low inaccessibility rates. At an inaccessibility rate of just over 0.08 or more, even with $TPR=TNR=0.7$, the performance is still better than the oblivious tool.  In most cases with inaccessibility rates above 0.1, even todays systems (with a $TNR=TPR=0.8-0.85$) still greatly outperform the oblivious tool, thereby offering a markedly improved experience (albeit an imperfect one). 

\begin{figure}[h]
  \centering
  \includegraphics[width=\linewidth]{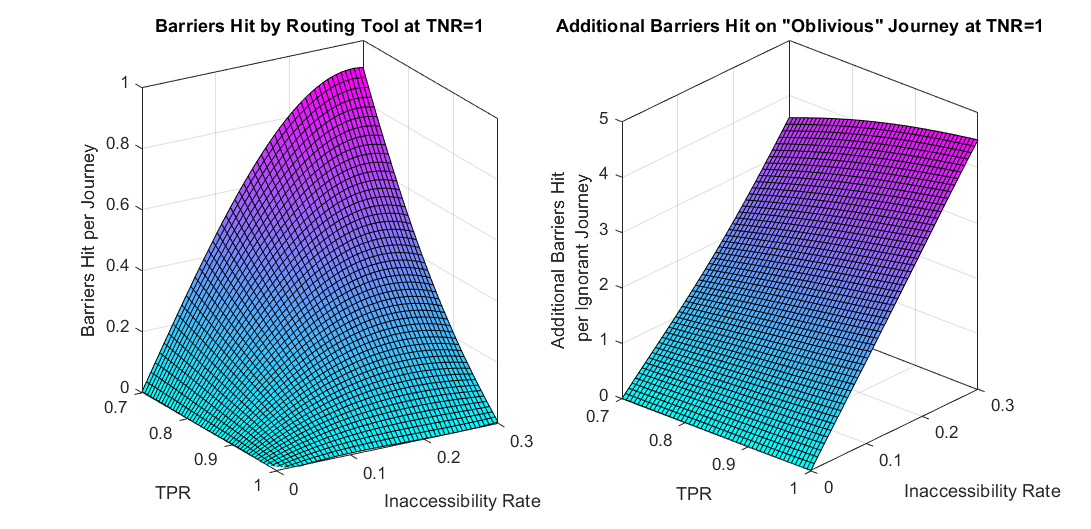}
  \caption{Illustration of average barriers encountered per journey under different conditions. On the left, Figure 8(i) illustrates the barriers encountered due to lower TPR’s (with TNR fixed to 1). On the right Figure 8(ii) illustrates the effect of different inaccessibility rates relative to an oblivious tool (i.e. where TPR=0). As can be seen, even a relatively poorly performing routing tool is an improvement on no-tool at all, but a relatively high TPR is required to avoid barriers on most occasions}
  \Description{}
\end{figure}
.
\begin{figure}[h]
  \centering
  \includegraphics[width=\linewidth]{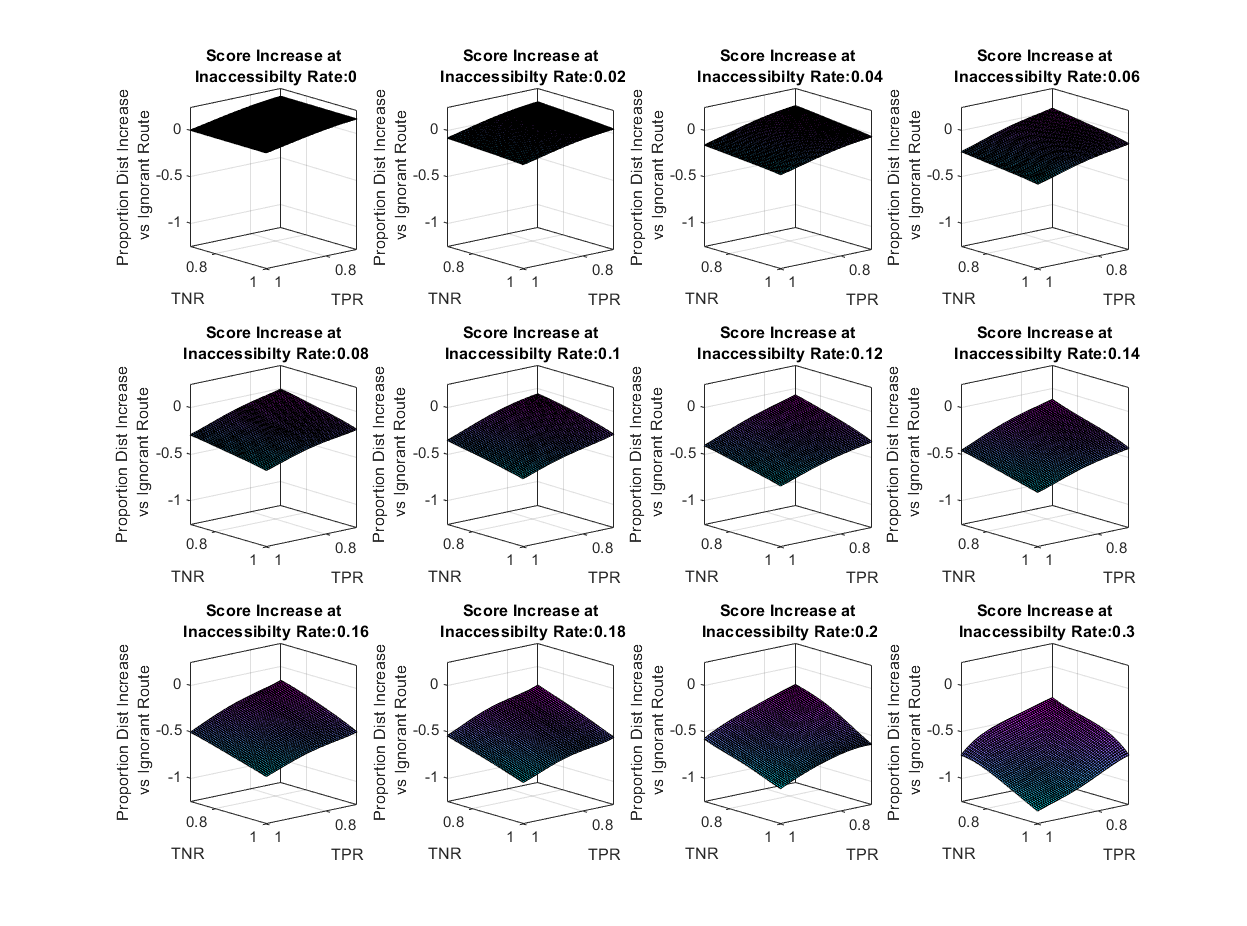}
  \caption{Performance Plots comparing the navigation tool to an oblivious tool with a TPR-0 (and TNR-1): negative scores are improvements compared to the ‘oblivious’ tool, and the lower the score the better. As can be seen, at very low inaccessibility rates, an oblivious tool performs slightly better than a tool with higher TPR and TNR (asides as very high TPR and TNR), however, this advantage quickly dissipates as soon as the inaccessibility rate somewhat increases. Indeed, at realistic levels of performance, an informed routing tool quickly becomes markedly improved upon an ‘oblivious’ tool. }
  \Description{}
\end{figure}

\section{Discussion}
\subsection{What is a navigation tool capable of?}
Our simulations give a clear indication of what can be achieved with a navigation-documentation tool, both with today’s performance, as well as likely future improvements in practice. The overall real-world effect of our simulations is as follows:

\textbf{Barrier Avoidance is not realistic in today’s built environment:} Our results show that barrier avoidance is unlikely to be a feasible approach towards providing a navigation tool, at least in today’s built environment. As such, the stated goal of barrier avoidance of many works in accessible navigation is practically impossible in today’s circumstances, even with a \textbf{perfect }documentation tool (and thus a fully informed navigation system). This is because even a moderately inaccessible city will lead to a significantly degraded navigation experience – a substantial proportion of journey’s will be \textbf{impossible}, and those journeys which remain possible will be a lot longer. For example, at what we expect is a relatively optimistic 20\% (or 0.2) inaccessibility rate\footnote{It is highly likely for some groups, such as electric wheelchair users, that the inaccessibility rate is over 0.3 in most cities. See \cite{Froehlich2020Towards}, which provides the  number of each type of barrier per 100m: in their best case city (Seattle), there was 0.3 obstacles per 100m, and 0.6 surface problems per 100m. Given the average segment length in Seattle was 71.61m, even half these values would produce an inaccessibility rate well beyond 0.3 (i.e. $(0.3+0.6)\times71.61/100 =0.64)$.}, this leads to 42.6\% of all journeys being inaccessible on average (i.e., there is no accessible route that can be taken at all), with an average journey distance increase of 15\% on those remaining routes which are navigable. Even at an inaccessibility rate of 10\% (an unlikely scenario), 13.25\% of journeys are impossible on average, and the average journey distance increase for the remaining journeys is 7.5\%, thus being far from being an ideal experience.

\textbf{Barrier Minimization and ‘quality aware’ navigation is feasible:} Against this context, we argue that the present goal of an accessibility navigation tool should be the more modest task of barrier minimization, rather than avoidance, which might well change if documentation becomes far more accurate (i.e., substantially above 0.95 for the TPR and TNR). Indeed, it is more realistic under normal circumstances – as can be seen, even a relatively poorly performing accessibility documentation tool can greatly improve upon an oblivious one, as long as the limitation is clear. An important implication for design flows from this: accessibility navigation tools should be expressly clear in that they are likely to be inaccurate, and promoted as reducing the chance accessibility barriers will be encountered. Once we design systems based on realistic expectations, we will have made tangible progress towards them being deployed in the day-to-day life of people with mobility impairments. 

\textbf{Better navigation would potentially be possible with a strategic effort to improve the built environment:} We assume that accessibility barriers are randomly distributed. However, there are a few cities who do make a concerted and strategic effort to provide accessible spaces in the right places, which may produce better navigation performance for a given inaccessibility rate. This strategy would align with work already done by city planners in respect of public transport services \cite{Ceder2016Public,Victor2012Urban} (and separately, the provision of cycling routes \cite{Winters2013Mapping,Larsen2013Build}), who carefully configure the spatial layout of routes to optimise performance, suggesting a related approach could potentially work with accessibility as well. For instance, it might be that if some accessible ‘trunk routes’ were to be deliberately built and identified in advance (perhaps in a grid formation as suggested for public transport in \cite{Ceder2016Public,Pemberton2020Optimising}), it would presumably be possible to improve substantially upon this performance, but in that case, there would be a lot of (expensive) building work required by a city or local council. Notably, the results of different approaches towards embedding accessibility in the built-environment can be simulated using our framework (by making the relevant edges always accessible and automatically detected by the documentation tool as being such) with a small modification, meaning these experiments can be done in the future.
\subsection{Implications for Human Experience}
The likely performance of a navigation tool across our range of scenarios set out in our results also has a range of important consequences on a human level, thereby leading to wider implications for the design of navigation systems for people with mobility impairments. 

\textbf{Navigation is not a panacea:} Our results show that even providing a perfect solution to the documentation of the built environment does not solve the navigation problem (although it would help in identifying barriers to be addressed). Instead, the built environment needs to be made a lot more accessible, in order to provide a reasonable experience for people with many mobility impairments. There is a real risk that the capability of tools to provide barrier minimization and thus a substantially better experience for many people with mobility impairments, leads to fewer physical improvements being made. For example, barrier minimization would likely lead to a lot better experience for manual wheelchair users (perhaps even a nearly equal one to people without mobility impairments) who would then be less concerned about physical improvements, but it would not address the problem for most electric wheelchair users (for whom physical improvements are necessary to give them a reasonable experience).  Given the complicated politics around disability and the built environment (see e.g. \cite{Kirkham2017WheelieMap:,Saha2021Urban}), the risk is that the most disabled people are disadvantaged by navigation systems, by less resources being spent on fixing the built environment, thus marginalizing them even further.

\textbf{Transparent and explainable navigation tools:} By acting as implicit performance metrics, our simulations can be used to manage expectations to provide better navigation systems: this works by being clear as to the likely navigation performance of a given system, which can be done provided there are estimates available of key variables (i.e. the inaccessibility rate, as well as the TPR / TNR of a documentation tool). Specifically, someone can be informed of the chance or probability of encountering one or more accessibility barriers – for example, it would be possible to say ‘\textit{On average you might encounter 0.3 accessibility barriers on your journey taking this route}’, or ‘\textit{Your journey will normally take between 15 and 25 minutes, considering the likelihood of having to re-route around accessibility barriers}’. Being transparent in this manner may provide a better experience for end users of these systems (e.g. journey planning can better consider the risk of being misdirected), as well as ensuring that these systems are used with realistic expectations in mind, because the expected performance under given circumstances can be directly quantified, especially in cities when the rough inaccessibility rate for different types of mobility impairments is known. 

Whilst the need for this has already been identified (see e.g. \cite{Tannert2019Analyzing}),  we are unaware of any widely used accessible navigation system that presents options and routes in terms of risk. The simulations we have presented allow for this risk to be estimated and presented to end users in the context of a navigation system, thus overcoming this hurdle if the accuracy of the documentation tool can be estimated, likewise with the distribution of accessibility barriers in a city. By being upfront with people with mobility impairments, we can expect that in return they would participate in the reporting of barriers, thus enabling a tool to become gradually more accurate and informative as it is used, thereby closing this important feedback loop. 

\textbf{Studying Navigation Tools:} Another implication of our results is that relatively small changes in inaccessibility rate can lead to markedly different user experiences (as the relationship is not linear). This has substantial implications for user experience studies of these tools. In particular, studies based on past experience (e.g. as with \cite{Gupta2020Towards,Hara2016design}) of using existing tools present this difficulty, because it would be difficult to separate out the performance of the tool from the level of accessibility in the built environment. In the immediate term, this may mean that these investigations should be done in respect of controlled simulations (rather than real world studies), so the performance of the tool and the variance across locations can be disaggregated from the interface design problem. These studies are particularly important going forward, to help determine which types of error should be avoided and for which groups of people. For instance, some people might be less concerned about a tool falsely reporting a journey is impossible, than by being sent on a journey that contains accessibility barriers. The right balance to be struck will likely vary substantially across different populations of people with mobility impairments, as well as being governed somewhat by individual preferences. 
\subsection{Expanding the role of simulations in Mobile Navigation}
The simulation approach that we have developed is potentially applicable in a wide range of circumstances, with some adaptions, making this work be of general import to the field of Mobile HCI beyond the accessible navigation. There are three main domains in which this work can be expanded.

\textbf{Other accessibility navigation problems:} The modelling approach that we have developed can also be applied to other accessibility related problems, with some modifications or different parameters. For example, a city planner could use this same approach to determine the impact of addressing specific accessibility barriers on the accessible navigation of their city (e.g. by setting purpose-built accessible edges/street segments to be always accessible and marked as such to a documentation tool, which would in turn be 100\% accurate in identifying it), thus optimizing the use of scarce public resources to make accessibility improvements. Similarly, our approach could also be applied to optimize the performance of a navigation tool (especially in terms of trade-offs between TNR and TPR) to provide the best experience for individual users, depending on their preferences and needs. 

With further expansion, it would also be possible to modify our approach to include user weights for different barriers: our simulations took the view that accessibility barriers should be always avoided (with the result of demonstrating the infeasibility of a barrier avoidance approach) and thus gave an infinite weight to an accessibility barrier, but this value can be changed. For instance, a manual wheelchair user might simply wish to somewhat reduce the amount of times they have to ‘curb hop’ due to a missing dropped curb, and would benefit from a system that gave a potential dropped curb a penalty (but a smaller one than automatically avoiding a given route) This would involve adding a distance weight (based on user preferences) for a perceived barrier of this nature and then picking the shortest route with this weight added. This approach can be expanded to include different weights for different types of barriers, depending on the person in question.  

\textbf{Navigation beyond accessibility:} Navigation in the real world has been a long-standing topic of interest in the wider MobileHCI community, including the design of routing interfaces. The minimization of the risk of error is one such concern that arises in a range of circumstances, be it as-the-crow-flies navigation \cite{Savino2020Point}, missing GPS information (where there is a degraded signal) \cite{Ranasinghe2019Visualising,Burigat2011Pedestrian}, or the provision of ‘scenic’ or more comfortable routes \cite{Johnson2017Beautiful}. These types of problems could also benefit from the identification of street elements that likely contain problems, be it a lack of a scenic feature, an area that is likely to have degraded mobile or GPS signals, or have other features which make navigation more challenging. Just as with accessibility barriers, there is likely to be imperfect information about these features, and thus a risk of encountering a suboptimal route. This is important for at least two reasons: first, a user can be given a choice of avoiding this risk, and secondly the risk of encountering a particular barrier can be quantified (which in turn could feed into interface design of a ‘quality aware’ navigation system \cite{Siriaraya2020Beyond}).  At the same time, the concept of an accessibility related feature can be a wide one if accessibility is considered in a holistic manner: for instance, some people seek to avoid areas of crime or poor weather \cite{Johnson2017Beautiful}, which may be sometimes be for reasons arising from a mental health condition, or another impairment, but is not generally considered as the primary concern. 

\textbf{Monte-Carlo Simulations as Practical and Implicit Performance Metrics:} Performance metrics are descriptive statistics used to summarise the result of a machine learning system, including a plethora of mobile sensing systems (especially activity trackers). Yet providing meaningful metrics is an important challenge, both for evaluating algorithms (by researchers) but also representing results to the public. Our simulation approach is important because it is distinct from the static and hard-edged performance metrics typically used for measuring classifier performance, and instead provides a practical measure of \textbf{real-world} performance focused on the \textbf{relevant problem} at hand. Given that accessibility documentation is similar to activity recognition problems (and indeed systems using inertial sensors and the motion of users to detect barriers are activity recognition systems) this is thus a new approach towards addressing the ‘performance metrics’ for activity recognition problem (see e.g.\cite{Hammerla2015Let's,Ward2011Performance} for a discussion of this problem). 

As such, this work demonstrates how to use simulations to construct\textbf{ implicit performance} metrics, that meaningfully translate static performance metrics into \textbf{problem-focused} ones. Whilst existing performance metrics are helpful when comparing the \textbf{relative} performance of algorithms, these simulations answer a different question: what performance is ‘good enough’ (in respect of a given \textbf{real-world} problem)? These implicit performance metrics are therefore an important complement to existing metrics traditionally used to measure classifier performance, and provide a platform for further exploration of real-world performance. This approach could in principle be adapted more widely across the domain of Mobile HCI, with Monte-Carlo simulations being used to provide information that is more relevant to end-users in respect of any system that depends on mobile-sensing. 
\subsection{Limitations}
Whilst our work clearly shows the benefits of using simulations to provide an indication of real-world performance, it should be observed that the simulations we have conducted are somewhat exploratory in nature. Our goal was to demonstrate some of the opportunities of following such an approach for the accessibility navigation/documentation problem, as well as illustrating how this can be practically implemented: this has been successful in raising a number of important concerns for the design of these systems and the advancement of this important research agenda. Nevertheless, given the complexity of the issue (and the simplifying assumptions we necessarily made), there are various parameter changes that potentially could lead to different results, including our choice of the bag of routes and city structures, our choice of navigation strategy, and our decision to constrain the length of routes, as well as smaller decisions in the design and implementation of our experiments. Our simulation assumed that the main objective was avoidance, with the effect that encounters with barriers were heavily penalized: some people with disabilities might be less demanding in their expectations (especially if their goal is to minimize discomfort). This means that there is perhaps an element of pessimism in our arguments, as a weaker tool might be useful for a minority of disabled people (even though this would be far away from the ambitions of the accessibility documentation agenda). Given the wide lacuna that this work fits into, we hope to see future investigations and simulations that explore alternative navigation methodologies and objectives.
\section{Conclusion}
This work illustrates the potential benefit of carefully designed computer simulations in advancing the agenda of accessibility documentation and navigation. In this work, we contributed a tractable approach for simulating a very large number of journeys in a reasonable time. We used this approach to demonstrate the performance boundaries placed on navigation tools for people with mobility impairments, thereby clearly illustrating the circumstances in which such tools can be useful and to what extent. The resulting implicit performance metrics led to clear design principles that respond to the boundaries placed on the performance of any such navigation tool, thus enabling an optimum approach towards the design of this specific type of navigation tool. Our simulations are also a supportive framework that enable the effective design of better navigation tools for people with mobility impairments and provide a means to evaluate their likely performance in a wide range of circumstances, thus enabling the management of the expectations of an end user. At the same time, with some adaption, this work can be used to design simulations that could address other types of navigation challenges, including those that are outside of the space of accessibility for people with mobility impairments.

\begin{acks}
This research was supported in part by the Monash eResearch Centre and eSolutions-Research Support Services through the use of the MonARCH HPC Cluster.
\end{acks}

\bibliographystyle{ACM-Reference-Format}
\bibliography{mhci}


\begin{thebibliography}{50}


\ifx \showCODEN    \undefined \def \showCODEN     #1{\unskip}     \fi
\ifx \showDOI      \undefined \def \showDOI       #1{#1}\fi
\ifx \showISBNx    \undefined \def \showISBNx     #1{\unskip}     \fi
\ifx \showISBNxiii \undefined \def \showISBNxiii  #1{\unskip}     \fi
\ifx \showISSN     \undefined \def \showISSN      #1{\unskip}     \fi
\ifx \showLCCN     \undefined \def \showLCCN      #1{\unskip}     \fi
\ifx \shownote     \undefined \def \shownote      #1{#1}          \fi
\ifx \showarticletitle \undefined \def \showarticletitle #1{#1}   \fi
\ifx \showURL      \undefined \def \showURL       {\relax}        \fi
\providecommand\bibfield[2]{#2}
\providecommand\bibinfo[2]{#2}
\providecommand\natexlab[1]{#1}
\providecommand\showeprint[2][]{arXiv:#2}

\bibitem[\protect\citeauthoryear{Bagenstos}{Bagenstos}{2020}]%
        {Bagenstos2020Towards}
\bibfield{author}{\bibinfo{person}{Samuel~R Bagenstos}.}
  \bibinfo{year}{2020}\natexlab{}.
\newblock \showarticletitle{Towards an Urban Disability Agenda}.
\newblock \bibinfo{journal}{\emph{Fordham Urban Law Journal}}
  \bibinfo{volume}{47}, \bibinfo{number}{5} (\bibinfo{year}{2020}),
  \bibinfo{pages}{1335--58}.
\newblock


\bibitem[\protect\citeauthoryear{Boeing}{Boeing}{2017}]%
        {Boeing2017OSMnx:}
\bibfield{author}{\bibinfo{person}{Geoff Boeing}.}
  \bibinfo{year}{2017}\natexlab{}.
\newblock \showarticletitle{OSMnx: New methods for acquiring, constructing,
  analyzing, and visualizing complex street networks}.
\newblock \bibinfo{journal}{\emph{Computers, Environment and Urban Systems}}
  \bibinfo{volume}{65} (\bibinfo{year}{2017}), \bibinfo{pages}{126–139}.
\newblock
\newblock
\shownote{publisher: Elsevier.}


\bibitem[\protect\citeauthoryear{Bowtell}{Bowtell}{2015}]%
        {Bowtell2015Assessing}
\bibfield{author}{\bibinfo{person}{James Bowtell}.}
  \bibinfo{year}{2015}\natexlab{}.
\newblock \showarticletitle{Assessing the value and market attractiveness of
  the accessible tourism industry in Europe: a focus on major travel and
  leisure companies}.
\newblock \bibinfo{journal}{\emph{Journal of Tourism Futures}}
  \bibinfo{volume}{1}, \bibinfo{number}{3} (\bibinfo{year}{2015}),
  \bibinfo{pages}{203–222}.
\newblock


\bibitem[\protect\citeauthoryear{Bromley, Matthews, and Thomas}{Bromley
  et~al\mbox{.}}{2007}]%
        {Bromley2007City}
\bibfield{author}{\bibinfo{person}{Rosemary~DF Bromley},
  \bibinfo{person}{David~L Matthews}, {and} \bibinfo{person}{Colin~J Thomas}.}
  \bibinfo{year}{2007}\natexlab{}.
\newblock \showarticletitle{City centre accessibility for wheelchair users: The
  consumer perspective and the planning implications}.
\newblock \bibinfo{journal}{\emph{Cities}} \bibinfo{volume}{24},
  \bibinfo{number}{3} (\bibinfo{year}{2007}), \bibinfo{pages}{229–241}.
\newblock
\newblock
\shownote{publisher: Elsevier.}


\bibitem[\protect\citeauthoryear{Burigat and Chittaro}{Burigat and
  Chittaro}{2011}]%
        {Burigat2011Pedestrian}
\bibfield{author}{\bibinfo{person}{Stefano Burigat} {and} \bibinfo{person}{Luca
  Chittaro}.} \bibinfo{year}{2011}\natexlab{}.
\newblock \showarticletitle{Pedestrian navigation with degraded GPS signal:
  investigating the effects of visualizing position uncertainty}.
\newblock \bibinfo{journal}{\emph{Proceedings of the 13th international
  conference on human computer interaction with mobile devices and services}},
  \bibinfo{pages}{221–230}.
\newblock


\bibitem[\protect\citeauthoryear{Ceder}{Ceder}{2016}]%
        {Ceder2016Public}
\bibfield{author}{\bibinfo{person}{Avishai Ceder}.}
  \bibinfo{year}{2016}\natexlab{}.
\newblock \bibinfo{booktitle}{\emph{Public transit planning and operation:
  Modeling, practice and behavior}}.
\newblock \bibinfo{publisher}{CRC press}.
\newblock


\bibitem[\protect\citeauthoryear{Church and Marston}{Church and
  Marston}{2003}]%
        {Church2003Measuring}
\bibfield{author}{\bibinfo{person}{Richard~L Church} {and}
  \bibinfo{person}{James~R Marston}.} \bibinfo{year}{2003}\natexlab{}.
\newblock \showarticletitle{Measuring accessibility for people with a
  disability}.
\newblock \bibinfo{journal}{\emph{Geographical Analysis}} \bibinfo{volume}{35},
  \bibinfo{number}{1} (\bibinfo{year}{2003}), \bibinfo{pages}{83–96}.
\newblock
\newblock
\shownote{publisher: Wiley Online Library.}


\bibitem[\protect\citeauthoryear{Ding, Parmanto, Karimi, Roongpiboonsopit,
  Pramana, Conahan, and Kasemsuppakorn}{Ding et~al\mbox{.}}{2007}]%
        {Ding2007Design}
\bibfield{author}{\bibinfo{person}{Dan Ding}, \bibinfo{person}{Bambang
  Parmanto}, \bibinfo{person}{Hassan~A Karimi}, \bibinfo{person}{Duangduen
  Roongpiboonsopit}, \bibinfo{person}{Gede Pramana}, \bibinfo{person}{Thomas
  Conahan}, {and} \bibinfo{person}{Piyawan Kasemsuppakorn}.}
  \bibinfo{year}{2007}\natexlab{}.
\newblock \showarticletitle{Design considerations for a personalized wheelchair
  navigation system}.
\newblock \bibinfo{journal}{\emph{2007 29th Annual International Conference of
  the IEEE Engineering in Medicine and Biology Society}},
  \bibinfo{pages}{4790–4793}.
\newblock


\bibitem[\protect\citeauthoryear{Froehlich, Brock, Caspi, Guerreiro, Hara,
  Kirkham, Schöning, and Tannert}{Froehlich et~al\mbox{.}}{2019}]%
        {Froehlich2019Grand}
\bibfield{author}{\bibinfo{person}{Jon~E Froehlich}, \bibinfo{person}{Anke
  Brock}, \bibinfo{person}{Anat Caspi}, \bibinfo{person}{João Guerreiro},
  \bibinfo{person}{Kotaro Hara}, \bibinfo{person}{Reuben Kirkham},
  \bibinfo{person}{Johannes Schöning}, {and} \bibinfo{person}{Benjamin
  Tannert}.} \bibinfo{year}{2019}\natexlab{}.
\newblock \showarticletitle{Grand challenges in accessible maps}.
\newblock \bibinfo{journal}{\emph{ACM Interactions}} \bibinfo{volume}{26},
  \bibinfo{number}{2} (\bibinfo{year}{2019}), \bibinfo{pages}{78--81}.
\newblock


\bibitem[\protect\citeauthoryear{Froehlich, Saugstad, Saha, and
  Johnson}{Froehlich et~al\mbox{.}}{2020}]%
        {Froehlich2020Towards}
\bibfield{author}{\bibinfo{person}{Jon~E Froehlich}, \bibinfo{person}{Michael
  Saugstad}, \bibinfo{person}{Manaswi Saha}, {and} \bibinfo{person}{Matthew
  Johnson}.} \bibinfo{year}{2020}\natexlab{}.
\newblock \showarticletitle{Towards Mapping and Assessing Sidewalk
  Accessibility Across Socio-cultural and Geographic Contexts}.
\newblock \bibinfo{journal}{\emph{Data4Good: Designing for Diversity and
  Development Workshop at AVI 2020,}} (\bibinfo{year}{2020}).
\newblock


\bibitem[\protect\citeauthoryear{Gani, Raychoudhury, Edinger, Mokrenko, Cao,
  and Zhang}{Gani et~al\mbox{.}}{2019}]%
        {Gani2019Smart}
\bibfield{author}{\bibinfo{person}{Md~Osman Gani}, \bibinfo{person}{Vaskar
  Raychoudhury}, \bibinfo{person}{Janick Edinger}, \bibinfo{person}{Valeria
  Mokrenko}, \bibinfo{person}{Zheng Cao}, {and} \bibinfo{person}{Ce Zhang}.}
  \bibinfo{year}{2019}\natexlab{}.
\newblock \showarticletitle{Smart surface classification for accessible routing
  through built environment: A crowd-sourced approach}.
\newblock \bibinfo{journal}{\emph{Proceedings of the 6th ACM International
  Conference on Systems for Energy-Efficient Buildings, Cities, and
  Transportation}}, \bibinfo{pages}{11–20}.
\newblock


\bibitem[\protect\citeauthoryear{Gupta, Abdolrahmani, Edwards, Cortez, Tumang,
  Majali, Lazaga, Tarra, Patil, Kuber, et~al\mbox{.}}{Gupta
  et~al\mbox{.}}{2020}]%
        {Gupta2020Towards}
\bibfield{author}{\bibinfo{person}{Maya Gupta}, \bibinfo{person}{Ali
  Abdolrahmani}, \bibinfo{person}{Emory Edwards}, \bibinfo{person}{Mayra
  Cortez}, \bibinfo{person}{Andrew Tumang}, \bibinfo{person}{Yasmin Majali},
  \bibinfo{person}{Marc Lazaga}, \bibinfo{person}{Samhitha Tarra},
  \bibinfo{person}{Prasad Patil}, \bibinfo{person}{Ravi Kuber},
  {et~al\mbox{.}}} \bibinfo{year}{2020}\natexlab{}.
\newblock \showarticletitle{Towards More Universal Wayfinding Technologies:
  Navigation Preferences Across Disabilities}.
\newblock \bibinfo{journal}{\emph{Proceedings of the 2020 CHI Conference on
  Human Factors in Computing Systems}}, \bibinfo{pages}{1–13}.
\newblock


\bibitem[\protect\citeauthoryear{Hammerla and Plötz}{Hammerla and
  Plötz}{2015}]%
        {Hammerla2015Let's}
\bibfield{author}{\bibinfo{person}{Nils~Y Hammerla} {and}
  \bibinfo{person}{Thomas Plötz}.} \bibinfo{year}{2015}\natexlab{}.
\newblock \showarticletitle{Let's (not) stick together: pairwise similarity
  biases cross-validation in activity recognition}.
\newblock \bibinfo{journal}{\emph{Ubicomp 2015}}, \bibinfo{pages}{1041–1051}.
\newblock


\bibitem[\protect\citeauthoryear{Hara, Chan, and Froehlich}{Hara
  et~al\mbox{.}}{2016}]%
        {Hara2016design}
\bibfield{author}{\bibinfo{person}{Kotaro Hara}, \bibinfo{person}{Christine
  Chan}, {and} \bibinfo{person}{Jon~E Froehlich}.}
  \bibinfo{year}{2016}\natexlab{}.
\newblock \showarticletitle{The design of assistive location-based technologies
  for people with ambulatory disabilities: A formative study}.
\newblock \bibinfo{journal}{\emph{Proceedings of the 2016 CHI Conference on
  Human Factors in Computing Systems}}, \bibinfo{pages}{1757–1768}.
\newblock


\bibitem[\protect\citeauthoryear{Hara, Sun, Moore, Jacobs, and Froehlich}{Hara
  et~al\mbox{.}}{2014}]%
        {Hara2014Tohme:}
\bibfield{author}{\bibinfo{person}{Kotaro Hara}, \bibinfo{person}{Jin Sun},
  \bibinfo{person}{Robert Moore}, \bibinfo{person}{David Jacobs}, {and}
  \bibinfo{person}{Jon Froehlich}.} \bibinfo{year}{2014}\natexlab{}.
\newblock \showarticletitle{Tohme: detecting curb ramps in google street view
  using crowdsourcing, computer vision, and machine learning}.
\newblock \bibinfo{journal}{\emph{Proceedings of the 27th annual ACM symposium
  on User interface software and technology}}, \bibinfo{pages}{189–204}.
\newblock


\bibitem[\protect\citeauthoryear{Harriehausen-Mühlbauer and
  Roth}{Harriehausen-Mühlbauer and Roth}{2016}]%
        {Harriehausen-Muhlbauer2016WheelScout-Barrier-Free}
\bibfield{author}{\bibinfo{person}{Bettina Harriehausen-Mühlbauer} {and}
  \bibinfo{person}{Jonas Roth}.} \bibinfo{year}{2016}\natexlab{}.
\newblock \showarticletitle{WheelScout-Barrier-Free Navigation}.
\newblock \bibinfo{journal}{\emph{Proceedings of SAI Intelligent Systems
  Conference}}, \bibinfo{pages}{1056–1063}.
\newblock


\bibitem[\protect\citeauthoryear{Iwasawa, Nagamine, Matsuo, and
  Eguchi~Yairi}{Iwasawa et~al\mbox{.}}{2015a}]%
        {Iwasawa2015Road}
\bibfield{author}{\bibinfo{person}{Yusuke Iwasawa}, \bibinfo{person}{Koya
  Nagamine}, \bibinfo{person}{Yutaka Matsuo}, {and} \bibinfo{person}{Ikuko
  Eguchi~Yairi}.} \bibinfo{year}{2015}\natexlab{a}.
\newblock \showarticletitle{Road sensing: Personal sensing and machine learning
  for development of large scale accessibility map}.
\newblock \bibinfo{journal}{\emph{Proceedings of the 17th International ACM
  SIGACCESS Conference on Computers \& Accessibility}},
  \bibinfo{pages}{335–336}.
\newblock


\bibitem[\protect\citeauthoryear{Iwasawa, Nagamine, Yairi, and Matsuo}{Iwasawa
  et~al\mbox{.}}{2015b}]%
        {Iwasawa2015Toward}
\bibfield{author}{\bibinfo{person}{Yusuke Iwasawa}, \bibinfo{person}{Kouya
  Nagamine}, \bibinfo{person}{Ikuko~Eguchi Yairi}, {and}
  \bibinfo{person}{Yutaka Matsuo}.} \bibinfo{year}{2015}\natexlab{b}.
\newblock \showarticletitle{Toward an automatic road accessibility information
  collecting and sharing based on human behavior sensing technologies of
  wheelchair users}.
\newblock \bibinfo{journal}{\emph{Procedia Computer Science}}
  \bibinfo{volume}{63} (\bibinfo{year}{2015}), \bibinfo{pages}{74–81}.
\newblock


\bibitem[\protect\citeauthoryear{Iwasawa and Yairi}{Iwasawa and Yairi}{2012}]%
        {Iwasawa2012Life-logging}
\bibfield{author}{\bibinfo{person}{Yusuke Iwasawa} {and}
  \bibinfo{person}{Ikuko~Eguchi Yairi}.} \bibinfo{year}{2012}\natexlab{}.
\newblock \showarticletitle{Life-logging of wheelchair driving on web maps for
  visualizing potential accidents and incidents}.
\newblock \bibinfo{journal}{\emph{Pacific Rim International Conference on
  Artificial Intelligence}}, \bibinfo{pages}{157–169}.
\newblock


\bibitem[\protect\citeauthoryear{Iwasawa, Yairi, and Matsuo}{Iwasawa
  et~al\mbox{.}}{2016}]%
        {Iwasawa2016Combining}
\bibfield{author}{\bibinfo{person}{Yusuke Iwasawa},
  \bibinfo{person}{Ikuko~Eguchi Yairi}, {and} \bibinfo{person}{Yutaka Matsuo}.}
  \bibinfo{year}{2016}\natexlab{}.
\newblock \showarticletitle{Combining Human Action Sensing of Wheelchair Users
  and Machine Learning for Autonomous Accessibility Data Collection}.
\newblock \bibinfo{journal}{\emph{IEICE Transactions on Information and
  Systems}} \bibinfo{volume}{99}, \bibinfo{number}{4} (\bibinfo{year}{2016}),
  \bibinfo{pages}{1153–1161}.
\newblock


\bibitem[\protect\citeauthoryear{Johnson, Henderson, Perry, Schöning, and
  Hecht}{Johnson et~al\mbox{.}}{2017}]%
        {Johnson2017Beautiful}
\bibfield{author}{\bibinfo{person}{Isaac Johnson}, \bibinfo{person}{Jessica
  Henderson}, \bibinfo{person}{Caitlin Perry}, \bibinfo{person}{Johannes
  Schöning}, {and} \bibinfo{person}{Brent Hecht}.}
  \bibinfo{year}{2017}\natexlab{}.
\newblock \showarticletitle{Beautiful… but at what cost? An examination of
  externalities in geographic vehicle routing}.
\newblock \bibinfo{journal}{\emph{Proceedings of the ACM on Interactive,
  Mobile, Wearable and Ubiquitous Technologies}} \bibinfo{volume}{1},
  \bibinfo{number}{2} (\bibinfo{year}{2017}), \bibinfo{pages}{1–21}.
\newblock
\newblock
\shownote{publisher: ACM New York, NY, USA.}


\bibitem[\protect\citeauthoryear{Kamaldin, Susan, Songwei, Chengkai, Liang,
  Saini, Hwee-Pink, and Hwee-Xian}{Kamaldin et~al\mbox{.}}{2019}]%
        {Kamaldin2019Smartbfa:}
\bibfield{author}{\bibinfo{person}{Nazir Kamaldin}, \bibinfo{person}{KEE
  Susan}, \bibinfo{person}{KONG Songwei}, \bibinfo{person}{LEE Chengkai},
  \bibinfo{person}{Huiguang Liang}, \bibinfo{person}{Alisha Saini},
  \bibinfo{person}{TAN Hwee-Pink}, {and} \bibinfo{person}{TAN Hwee-Xian}.}
  \bibinfo{year}{2019}\natexlab{}.
\newblock \showarticletitle{Smartbfa: A passive crowdsourcing system for
  point-to-point barrier-free access}.
\newblock \bibinfo{journal}{\emph{2019 IEEE 44th Conference on Local Computer
  Networks (LCN)}}, \bibinfo{pages}{34–41}.
\newblock


\bibitem[\protect\citeauthoryear{Kirkham, Ebassa, Montague, Morrissey,
  Vlachokyriakos, Weise, and Olivier}{Kirkham et~al\mbox{.}}{2017}]%
        {Kirkham2017WheelieMap:}
\bibfield{author}{\bibinfo{person}{Reuben Kirkham}, \bibinfo{person}{Romeo
  Ebassa}, \bibinfo{person}{Kyle Montague}, \bibinfo{person}{Kellie Morrissey},
  \bibinfo{person}{Vasilis Vlachokyriakos}, \bibinfo{person}{Sebastian Weise},
  {and} \bibinfo{person}{Patrick Olivier}.} \bibinfo{year}{2017}\natexlab{}.
\newblock \showarticletitle{WheelieMap: An Exploratory System for Qualitative
  Reports of Inaccessibility in the Built Environment}.
\newblock \bibinfo{journal}{\emph{Proceedings of the 19th International
  Conference on Human-Computer Interaction with Mobile Devices and Services}},
  \bibinfo{pages}{38}.
\newblock


\bibitem[\protect\citeauthoryear{Kristensson, Lilley, Black, and
  Waller}{Kristensson et~al\mbox{.}}{2020}]%
        {Kristensson2020Design}
\bibfield{author}{\bibinfo{person}{Per~Ola Kristensson}, \bibinfo{person}{James
  Lilley}, \bibinfo{person}{Rolf Black}, {and} \bibinfo{person}{Annalu
  Waller}.} \bibinfo{year}{2020}\natexlab{}.
\newblock \showarticletitle{A Design Engineering Approach for Quantitatively
  Exploring Context-Aware Sentence Retrieval for Nonspeaking Individuals with
  Motor Disabilities}.
\newblock \bibinfo{journal}{\emph{Proceedings of the 2020 CHI Conference on
  Human Factors in Computing Systems}}, \bibinfo{pages}{1–11}.
\newblock
\showISBNx{978-1-4503-6708-0}
\urldef\tempurl%
\url{https://doi.org/10.1145/3313831.3376525}
\showDOI{\tempurl}
\newblock
\shownote{event-place: Honolulu, HI, USA.}


\bibitem[\protect\citeauthoryear{Kurauchi, Abe, Konishi, and Seshimo}{Kurauchi
  et~al\mbox{.}}{2019}]%
        {Kurauchi2019Barrier}
\bibfield{author}{\bibinfo{person}{Yuki Kurauchi}, \bibinfo{person}{Naoto Abe},
  \bibinfo{person}{Hiroshi Konishi}, {and} \bibinfo{person}{Hitoshi Seshimo}.}
  \bibinfo{year}{2019}\natexlab{}.
\newblock \showarticletitle{Barrier Detection Using Sensor Data from Multiple
  Modes of Transportation with Data Augmentation}.
\newblock \bibinfo{journal}{\emph{2019 IEEE 43rd Annual Computer Software and
  Applications Conference (COMPSAC)}}  \bibinfo{volume}{1},
  \bibinfo{pages}{667–675}.
\newblock


\bibitem[\protect\citeauthoryear{Lange, Tannert, and Kirkham}{Lange
  et~al\mbox{.}}{2021}]%
        {Lange2021}
\bibfield{author}{\bibinfo{person}{Marvin Lange}, \bibinfo{person}{Benjamin
  Tannert}, {and} \bibinfo{person}{Reuben Kirkham}.}
  \bibinfo{year}{2021}\natexlab{}.
\newblock \showarticletitle{Strategically using Applied Machine Learning for
  Accessibility Documentation in the Built Environment}.
\newblock \bibinfo{journal}{\emph{Interact 2021}}.
\newblock


\bibitem[\protect\citeauthoryear{Larsen, Patterson, and El-Geneidy}{Larsen
  et~al\mbox{.}}{2013}]%
        {Larsen2013Build}
\bibfield{author}{\bibinfo{person}{Jacob Larsen}, \bibinfo{person}{Zachary
  Patterson}, {and} \bibinfo{person}{Ahmed El-Geneidy}.}
  \bibinfo{year}{2013}\natexlab{}.
\newblock \showarticletitle{Build it. But where? The use of geographic
  information systems in identifying locations for new cycling infrastructure}.
\newblock \bibinfo{journal}{\emph{International Journal of Sustainable
  Transportation}} \bibinfo{volume}{7}, \bibinfo{number}{4}
  (\bibinfo{year}{2013}), \bibinfo{pages}{299–317}.
\newblock
\newblock
\shownote{publisher: Taylor \& Francis.}


\bibitem[\protect\citeauthoryear{Mascetti, Civitarese, El~Malak, and
  Bettini}{Mascetti et~al\mbox{.}}{2020}]%
        {Mascetti2020SmartWheels:}
\bibfield{author}{\bibinfo{person}{Sergio Mascetti}, \bibinfo{person}{Gabriele
  Civitarese}, \bibinfo{person}{Omar El~Malak}, {and} \bibinfo{person}{Claudio
  Bettini}.} \bibinfo{year}{2020}\natexlab{}.
\newblock \showarticletitle{SmartWheels: Detecting urban features for
  wheelchair users’ navigation}.
\newblock \bibinfo{journal}{\emph{Pervasive and Mobile Computing}}
  (\bibinfo{year}{2020}), \bibinfo{pages}{101115}.
\newblock


\bibitem[\protect\citeauthoryear{Morris}{Morris}{2020}]%
        {Morris2020AI}
\bibfield{author}{\bibinfo{person}{Meredith~Ringel Morris}.}
  \bibinfo{year}{2020}\natexlab{}.
\newblock \showarticletitle{AI and accessibility}.
\newblock \bibinfo{journal}{\emph{Commun. ACM}} \bibinfo{volume}{63},
  \bibinfo{number}{6} (\bibinfo{year}{2020}), \bibinfo{pages}{35–37}.
\newblock
\newblock
\shownote{publisher: ACM New York, NY, USA.}


\bibitem[\protect\citeauthoryear{Mourcou, Fleury, Dupuy, Diot, Franco, and
  Vuillerme}{Mourcou et~al\mbox{.}}{2013}]%
        {Mourcou2013Wegoto:}
\bibfield{author}{\bibinfo{person}{Quentin Mourcou}, \bibinfo{person}{Anthony
  Fleury}, \bibinfo{person}{Pascal Dupuy}, \bibinfo{person}{B Diot},
  \bibinfo{person}{C Franco}, {and} \bibinfo{person}{Nicolas Vuillerme}.}
  \bibinfo{year}{2013}\natexlab{}.
\newblock \showarticletitle{Wegoto: A Smartphone-based approach to assess and
  improve accessibility for wheelchair users}.
\newblock \bibinfo{journal}{\emph{IEEE EMBS 2013}},
  \bibinfo{pages}{1194–1197}.
\newblock


\bibitem[\protect\citeauthoryear{Organization et~al\mbox{.}}{Organization
  et~al\mbox{.}}{2011}]%
        {Organization2011World}
\bibfield{author}{\bibinfo{person}{World~Health Organization} {et~al\mbox{.}}}
  \bibinfo{year}{2011}\natexlab{}.
\newblock \bibinfo{booktitle}{\emph{World report on disability 2011}}.
\newblock \bibinfo{publisher}{World Health Organization}.
\newblock


\bibitem[\protect\citeauthoryear{Pemberton}{Pemberton}{2020}]%
        {Pemberton2020Optimising}
\bibfield{author}{\bibinfo{person}{Steve Pemberton}.}
  \bibinfo{year}{2020}\natexlab{}.
\newblock \showarticletitle{Optimising Melbourne's bus routes for real-life
  travel patterns}.
\newblock \bibinfo{journal}{\emph{Case Studies on Transport Policy}}
  \bibinfo{volume}{8}, \bibinfo{number}{3} (\bibinfo{year}{2020}),
  \bibinfo{pages}{1038–1052}.
\newblock
\newblock
\shownote{publisher: Elsevier.}


\bibitem[\protect\citeauthoryear{Poppe, Rienks, and van Dijk}{Poppe
  et~al\mbox{.}}{2007}]%
        {Poppe2007Evaluating}
\bibfield{author}{\bibinfo{person}{Ronald Poppe}, \bibinfo{person}{Rutger
  Rienks}, {and} \bibinfo{person}{Betsy van Dijk}.}
  \bibinfo{year}{2007}\natexlab{}.
\newblock \showarticletitle{Evaluating the future of HCI: challenges for the
  evaluation of emerging applications}.
\newblock In \bibinfo{booktitle}{\emph{Artificial Intelligence for Human
  Computing}}. \bibinfo{publisher}{Springer}, \bibinfo{pages}{234–250}.
\newblock


\bibitem[\protect\citeauthoryear{Ranasinghe, Schiestel, and Kray}{Ranasinghe
  et~al\mbox{.}}{2019}]%
        {Ranasinghe2019Visualising}
\bibfield{author}{\bibinfo{person}{Champika Ranasinghe},
  \bibinfo{person}{Nicholas Schiestel}, {and} \bibinfo{person}{Christian
  Kray}.} \bibinfo{year}{2019}\natexlab{}.
\newblock \showarticletitle{Visualising location uncertainty to support
  navigation under degraded gps signals: A comparison study}.
\newblock \bibinfo{journal}{\emph{Proceedings of the 21st International
  Conference on Human-Computer Interaction with Mobile Devices and Services}},
  \bibinfo{pages}{1–11}.
\newblock


\bibitem[\protect\citeauthoryear{Rice, Jacobson, Pfoser, Curtin, Qin, Coll,
  Rice, Paez, and Aburizaiza}{Rice et~al\mbox{.}}{2018}]%
        {Rice2018Quality}
\bibfield{author}{\bibinfo{person}{Matthew~T Rice}, \bibinfo{person}{Dan
  Jacobson}, \bibinfo{person}{Dieter Pfoser}, \bibinfo{person}{Kevin~M Curtin},
  \bibinfo{person}{Han Qin}, \bibinfo{person}{Kerry Coll},
  \bibinfo{person}{Rebecca Rice}, \bibinfo{person}{Fabiana Paez}, {and}
  \bibinfo{person}{Ahmad~Omar Aburizaiza}.} \bibinfo{year}{2018}\natexlab{}.
\newblock \showarticletitle{Quality Assessment and Accessibility Mapping in an
  Image-Based Geocrowdsourcing Testbed}.
\newblock \bibinfo{journal}{\emph{Cartographica: The International Journal for
  Geographic Information and Geovisualization}} \bibinfo{volume}{53},
  \bibinfo{number}{1} (\bibinfo{year}{2018}), \bibinfo{pages}{1–14}.
\newblock


\bibitem[\protect\citeauthoryear{Saha, Chauhan, Patil, Kangas, Heer, and
  Froehlich}{Saha et~al\mbox{.}}{2021}]%
        {Saha2021Urban}
\bibfield{author}{\bibinfo{person}{Manaswi Saha}, \bibinfo{person}{Devanshi
  Chauhan}, \bibinfo{person}{Siddhant Patil}, \bibinfo{person}{Rachel Kangas},
  \bibinfo{person}{Jeffrey Heer}, {and} \bibinfo{person}{Jon~E Froehlich}.}
  \bibinfo{year}{2021}\natexlab{}.
\newblock \showarticletitle{Urban Accessibility as a Socio-Political Problem: A
  Multi-Stakeholder Analysis}.
\newblock \bibinfo{journal}{\emph{Proceedings of the ACM on Human-Computer
  Interaction}} \bibinfo{volume}{4}, \bibinfo{number}{CSCW3}
  (\bibinfo{year}{2021}), \bibinfo{pages}{1–26}.
\newblock
\newblock
\shownote{publisher: ACM New York, NY, USA.}


\bibitem[\protect\citeauthoryear{Saha, Saugstad, Maddali, Zeng, Holland, Bower,
  Dash, Chen, Li, Hara, et~al\mbox{.}}{Saha et~al\mbox{.}}{2019}]%
        {Saha2019Project}
\bibfield{author}{\bibinfo{person}{Manaswi Saha}, \bibinfo{person}{Michael
  Saugstad}, \bibinfo{person}{Hanuma~Teja Maddali}, \bibinfo{person}{Aileen
  Zeng}, \bibinfo{person}{Ryan Holland}, \bibinfo{person}{Steven Bower},
  \bibinfo{person}{Aditya Dash}, \bibinfo{person}{Sage Chen},
  \bibinfo{person}{Anthony Li}, \bibinfo{person}{Kotaro Hara}, {et~al\mbox{.}}}
  \bibinfo{year}{2019}\natexlab{}.
\newblock \showarticletitle{Project sidewalk: A web-based crowdsourcing tool
  for collecting sidewalk accessibility data at scale}.
\newblock \bibinfo{journal}{\emph{Proceedings of the 2019 CHI Conference on
  Human Factors in Computing Systems}}, \bibinfo{pages}{1–14}.
\newblock


\bibitem[\protect\citeauthoryear{Savino, Meyer, Schade, Tenbrink, and
  Schöning}{Savino et~al\mbox{.}}{2020}]%
        {Savino2020Point}
\bibfield{author}{\bibinfo{person}{Gian-Luca Savino}, \bibinfo{person}{Laura
  Meyer}, \bibinfo{person}{Eve Emily~Sophie Schade}, \bibinfo{person}{Thora
  Tenbrink}, {and} \bibinfo{person}{Johannes Schöning}.}
  \bibinfo{year}{2020}\natexlab{}.
\newblock \showarticletitle{Point Me In the Right Direction: Understanding User
  Behaviour with As-The-Crow-Flies Navigation}.
\newblock \bibinfo{journal}{\emph{22nd International Conference on
  Human-Computer Interaction with Mobile Devices and Services}},
  \bibinfo{pages}{1–11}.
\newblock


\bibitem[\protect\citeauthoryear{Siriaraya, Wang, Zhang, Wakamiya, Jeszenszky,
  Kawai, and Jatowt}{Siriaraya et~al\mbox{.}}{2020}]%
        {Siriaraya2020Beyond}
\bibfield{author}{\bibinfo{person}{Panote Siriaraya}, \bibinfo{person}{Yuanyuan
  Wang}, \bibinfo{person}{Yihong Zhang}, \bibinfo{person}{Shoko Wakamiya},
  \bibinfo{person}{Péter Jeszenszky}, \bibinfo{person}{Yukiko Kawai}, {and}
  \bibinfo{person}{Adam Jatowt}.} \bibinfo{year}{2020}\natexlab{}.
\newblock \showarticletitle{Beyond the Shortest Route: A Survey on
  Quality-Aware Route Navigation for Pedestrians}.
\newblock \bibinfo{journal}{\emph{IEEE Access}}  \bibinfo{volume}{8}
  (\bibinfo{year}{2020}), \bibinfo{pages}{135569–135590}.
\newblock
\newblock
\shownote{publisher: IEEE.}


\bibitem[\protect\citeauthoryear{Tannert, Kirkham, and Schöning}{Tannert
  et~al\mbox{.}}{2019}]%
        {Tannert2019Analyzing}
\bibfield{author}{\bibinfo{person}{Benjamin Tannert}, \bibinfo{person}{Reuben
  Kirkham}, {and} \bibinfo{person}{Johannes Schöning}.}
  \bibinfo{year}{2019}\natexlab{}.
\newblock \showarticletitle{Analyzing Accessibility Barriers Using Cost-Benefit
  Analysis to Design Reliable Navigation Services for Wheelchair Users.}
\newblock \bibinfo{journal}{\emph{Interact 2019}}.
\newblock


\bibitem[\protect\citeauthoryear{Velho, Holloway, Symonds, and Balmer}{Velho
  et~al\mbox{.}}{2016}]%
        {Velho2016effect}
\bibfield{author}{\bibinfo{person}{Raquel Velho}, \bibinfo{person}{Catherine
  Holloway}, \bibinfo{person}{Andrew Symonds}, {and} \bibinfo{person}{Brian
  Balmer}.} \bibinfo{year}{2016}\natexlab{}.
\newblock \showarticletitle{The effect of transport accessibility on the social
  inclusion of wheelchair users: A mixed method analysis}.
\newblock \bibinfo{journal}{\emph{Social Inclusion}} \bibinfo{volume}{4},
  \bibinfo{number}{3} (\bibinfo{year}{2016}), \bibinfo{pages}{24–35}.
\newblock


\bibitem[\protect\citeauthoryear{Victor and Ponnuswamy}{Victor and
  Ponnuswamy}{2012}]%
        {Victor2012Urban}
\bibfield{author}{\bibinfo{person}{D~Johnson Victor} {and} \bibinfo{person}{S
  Ponnuswamy}.} \bibinfo{year}{2012}\natexlab{}.
\newblock \bibinfo{booktitle}{\emph{Urban transportation: planning, operation
  and management}}.
\newblock \bibinfo{publisher}{Tata McGraw-Hill Education}.
\newblock


\bibitem[\protect\citeauthoryear{Völkel, Kühn, and Weber}{Völkel
  et~al\mbox{.}}{2008}]%
        {Volkel2008Mobility}
\bibfield{author}{\bibinfo{person}{Thorsten Völkel}, \bibinfo{person}{Romina
  Kühn}, {and} \bibinfo{person}{Gerhard Weber}.}
  \bibinfo{year}{2008}\natexlab{}.
\newblock \showarticletitle{Mobility impaired pedestrians are not cars:
  Requirements for the annotation of geographical data}.
\newblock \bibinfo{journal}{\emph{International Conference on Computers for
  Handicapped Persons}}, \bibinfo{pages}{1085–1092}.
\newblock


\bibitem[\protect\citeauthoryear{Ward, Lukowicz, and Gellersen}{Ward
  et~al\mbox{.}}{2011}]%
        {Ward2011Performance}
\bibfield{author}{\bibinfo{person}{Jamie~A Ward}, \bibinfo{person}{Paul
  Lukowicz}, {and} \bibinfo{person}{Hans~W Gellersen}.}
  \bibinfo{year}{2011}\natexlab{}.
\newblock \showarticletitle{Performance metrics for activity recognition}.
\newblock \bibinfo{journal}{\emph{ACM Transactions on Intelligent Systems and
  Technology (TIST)}} \bibinfo{volume}{2}, \bibinfo{number}{1}
  (\bibinfo{year}{2011}), \bibinfo{pages}{6}.
\newblock


\bibitem[\protect\citeauthoryear{Weld, Jang, Li, Zeng, Heimerl, and
  Froehlich}{Weld et~al\mbox{.}}{2019}]%
        {Weld2019Deep}
\bibfield{author}{\bibinfo{person}{Galen Weld}, \bibinfo{person}{Esther Jang},
  \bibinfo{person}{Anthony Li}, \bibinfo{person}{Aileen Zeng},
  \bibinfo{person}{Kurtis Heimerl}, {and} \bibinfo{person}{Jon~E Froehlich}.}
  \bibinfo{year}{2019}\natexlab{}.
\newblock \showarticletitle{Deep Learning for Automatically Detecting Sidewalk
  Accessibility Problems Using Streetscape Imagery}.
\newblock \bibinfo{journal}{\emph{The 21st International ACM SIGACCESS
  Conference on Computers and Accessibility}}, \bibinfo{pages}{196–209}.
\newblock


\bibitem[\protect\citeauthoryear{Whittaker, Alper, Bennett, Hendren, Kaziunas,
  Mills, Morris, Rankin, Rogers, Salas, et~al\mbox{.}}{Whittaker
  et~al\mbox{.}}{2019}]%
        {Whittaker2019Disability}
\bibfield{author}{\bibinfo{person}{Meredith Whittaker}, \bibinfo{person}{Meryl
  Alper}, \bibinfo{person}{Cynthia~L Bennett}, \bibinfo{person}{Sara Hendren},
  \bibinfo{person}{Liz Kaziunas}, \bibinfo{person}{Mara Mills},
  \bibinfo{person}{Meredith~Ringel Morris}, \bibinfo{person}{Joy Rankin},
  \bibinfo{person}{Emily Rogers}, \bibinfo{person}{Marcel Salas},
  {et~al\mbox{.}}} \bibinfo{year}{2019}\natexlab{}.
\newblock \showarticletitle{Disability, Bias, and AI}.
\newblock \bibinfo{journal}{\emph{AI Now Institute, November}}
  (\bibinfo{year}{2019}).
\newblock


\bibitem[\protect\citeauthoryear{Winters, Brauer, Setton, and Teschke}{Winters
  et~al\mbox{.}}{2013}]%
        {Winters2013Mapping}
\bibfield{author}{\bibinfo{person}{Meghan Winters}, \bibinfo{person}{Michael
  Brauer}, \bibinfo{person}{Eleanor~M Setton}, {and} \bibinfo{person}{Kay
  Teschke}.} \bibinfo{year}{2013}\natexlab{}.
\newblock \showarticletitle{Mapping bikeability: a spatial tool to support
  sustainable travel}.
\newblock \bibinfo{journal}{\emph{Environment and Planning B: Planning and
  Design}} \bibinfo{volume}{40}, \bibinfo{number}{5} (\bibinfo{year}{2013}),
  \bibinfo{pages}{865–883}.
\newblock
\newblock
\shownote{publisher: SAGE Publications Sage UK: London, England.}


\bibitem[\protect\citeauthoryear{Wu, Liu, and Yuan}{Wu et~al\mbox{.}}{2020}]%
        {Wu2020mobile-based}
\bibfield{author}{\bibinfo{person}{Yenchun~Jim Wu}, \bibinfo{person}{Wan-Ju
  Liu}, {and} \bibinfo{person}{Chih-Hung Yuan}.}
  \bibinfo{year}{2020}\natexlab{}.
\newblock \showarticletitle{A mobile-based barrier-free service transportation
  platform for people with disabilities}.
\newblock \bibinfo{journal}{\emph{Computers in Human Behavior}}
  \bibinfo{volume}{107} (\bibinfo{year}{2020}), \bibinfo{pages}{105776}.
\newblock
\newblock
\shownote{publisher: Elsevier.}


\bibitem[\protect\citeauthoryear{Yairi, Takahashi, Watanabe, Nagamine,
  Fukushima, Matsuo, and Iwasawa}{Yairi et~al\mbox{.}}{2019}]%
        {Yairi2019Estimating}
\bibfield{author}{\bibinfo{person}{Ikuko~Eguchi Yairi}, \bibinfo{person}{Hiroki
  Takahashi}, \bibinfo{person}{Takumi Watanabe}, \bibinfo{person}{Kouya
  Nagamine}, \bibinfo{person}{Yusuke Fukushima}, \bibinfo{person}{Yutaka
  Matsuo}, {and} \bibinfo{person}{Yusuke Iwasawa}.}
  \bibinfo{year}{2019}\natexlab{}.
\newblock \showarticletitle{Estimating Spatiotemporal Information from
  Behavioral Sensing Data of Wheelchair Users by Machine Learning
  Technologies}.
\newblock \bibinfo{journal}{\emph{Information}} \bibinfo{volume}{10},
  \bibinfo{number}{3} (\bibinfo{year}{2019}), \bibinfo{pages}{114}.
\newblock


\bibitem[\protect\citeauthoryear{Yang and Diez-Roux}{Yang and
  Diez-Roux}{2012}]%
        {Yang2012Walking}
\bibfield{author}{\bibinfo{person}{Yong Yang} {and} \bibinfo{person}{Ana~V
  Diez-Roux}.} \bibinfo{year}{2012}\natexlab{}.
\newblock \showarticletitle{Walking distance by trip purpose and population
  subgroups}.
\newblock \bibinfo{journal}{\emph{American journal of preventive medicine}}
  \bibinfo{volume}{43}, \bibinfo{number}{1} (\bibinfo{year}{2012}),
  \bibinfo{pages}{11–19}.
\newblock
\newblock
\shownote{publisher: Elsevier.}


\end{thebibliography}


\end{document}